\documentclass[pra,twocolumn,showpacs,preprintnumbers,amsmath,amssymb]{revtex4}

\usepackage{graphicx}
\usepackage{dcolumn}
\usepackage{bm}

\usepackage[T1]{fontenc}
\usepackage[latin1]{inputenc}
\usepackage[english]{babel}
\usepackage{amsmath}
\usepackage{amssymb}
\usepackage{amstext}
\usepackage{bbold}
\usepackage{braket}
\usepackage{relsize}
\usepackage{subfigure}
\usepackage{booktabs}
\usepackage{natbib}

\newcommand{\vs}[1]{\ensuremath{\boldsymbol{#1}}}
\newcommand{\vect}[1]{\ensuremath{\bm{#1}}}
\newcommand{\eq}[1]{\begin{equation} #1 \end{equation}}
\newcommand{\eqa}[1]{\begin{eqnarray} #1 \end{eqnarray}}

\newcommand{\be}{\begin{equation}}
\newcommand{\ee}{\end{equation}}
\newcommand{\ba}{\begin{eqnarray}}
\newcommand{\ea}{\end{eqnarray}}

\newcommand{\ue}{\mathrm{e}}

\newcommand{\bA}{{\bm A}}

\newcommand{\br}{{\bm r}}
\newcommand{\ri}{\mathrm{i}}

\begin{document}

\title{
Quantum disorder in the spatially completely anisotropic triangular lattice \newline
II: frustrated hard-core bosons
}
\author{Philipp Hauke}
    \email{philipp.hauke@icfo.es}
    \affiliation{ICFO -- Institut de Ci\`{e}ncies Fot\`{o}niques, Parc Mediterrani de la Tecnologia, 08860 Castelldefels, Spain}
    \affiliation{Institute for Quantum Optics and Quantum Information of the Austrian Academy of Sciences, 6020 Innsbruck, Austria}

\date{\today}

\begin{abstract}
Spin liquids occuring in 2D frustrated spin systems were initially assumed to appear at strongest frustration, but evidence grows that they more likely intervene at transitions between two different types of order. 
To identify if this is more general, we here analyze a generalization of the spatially anisotropic triangular lattice (SATL) with antiferromagnetic XY interactions, the spatially \emph{completely} anisotropic triangular lattice (SCATL). 
This model can be implemented in experiments with trapped ions, ultra-small Josephson junctions, or ultracold atoms in optical lattices. 
Using Takahashi's modified spin-wave theory, we find indications that indeed two different kinds of order are always separated by phases without magnetic long-range order. 
Our results further suggest that two gapped, magnetically-disordered phases, identified as distinct in the SATL, are actually continuously connected via the additional anisotropy of the SCATL. 
As these results indicate, this additional anisotropy -- allowing to approach quantum-disordered phases from different angles -- can give fundamental insight into the nature of quantum disordered phases.
We complement our results by exact diagonalizations, which also indicate that in part of the gapped non-magnetic phase, chiral long-range correlations could survive. 
\end{abstract}


\maketitle

\section{Introduction}

In recent years, the implementation of strongly-correlated lattice boson models in optical-lattice experiments has seen tremendous advances (see, e.g., \cite{Lewenstein2012} and references therein). Particularly appealing in this respect are models with frustration, 
arising for instance from the coupling of the bosons to an artificial magnetic field \cite{Jaksch2003, Sorensen2005,Lin2009b,Aidelsburger2011,Jimenez2012}, or from a periodical shaking of the optical lattice \cite{Eckardt2010,Struck2011a,Sacha2011,Struck2012,Hauke2012c}. 
Frustration in the intersite hopping is formally equivalent to a description of the system in a rotating reference frame, which implies that the system is subject to the spontaneous appearance of vortices. 
Such vortices can form ordered arrays (vortex crystals) coexisting with Bose condensation, which consequently takes place in a macroscopic wavefunction sustaining persisting circulating currents (see Ref.~\cite{Goldbaum2008} and references therein); 
or they can even disrupt condensation completely, and lead to a disordered insulating state \cite{Garcia-Ripoll2007}. 
Such disordered states are notoriously difficult to study theoretically.

In the limit of strong on-site repulsion, such bosonic models can be exactly mapped to $S=1/2$ XY antiferromagnets \cite{Diep2004}.
In two dimensions, such frustrated XY models exhibit ground states with spiral order, representing the magnetic counterpart to the aforementioned Bose-condensed states with vortex arrays. 
More strikingly, the interplay between quantum fluctuations and frustration may lead to disordered spin-liquid states, which are in one-to-one correspondence with bosonic insulating phases.  
While our main motivation stems from the recent advances in ultracold atoms in optical lattices, antiferromagnetic (AFM) XY models are relevant to a number of systems: they can be regarded as the limiting case of AFM Hamiltonians with planar anisotropy in the couplings, which describe frustrated AFM materials, and they govern the physics of planarly trapped ions loaded into an optical lattice \cite{Schmied2008} or Cooper pairs in arrays of ultra-small Josephson junctions \cite{Fazio2001}.

Planar systems of bosons in optical lattices can be described by the Bose--Hubbard Hamiltonian
\begin{equation}
{\cal H}_{\mathrm{BH}} = \sum_{\braket{i,j}} \frac{t_{ij}}{2} \left(b_i^{\dagger} b_j + \text{h.c.}\right) + \frac{U}{2} \sum_i n_i (n_i - 1) 
\label{eq:BH}
\end{equation}
where $b_i$, $b_i^{\dagger}$ are bosonic operators, $n_i = b_i^{\dagger} b_i$, and $\braket{i,j}$ represents pairs of nearest-neighbor (NN) sites. 
We consider AFM hopping amplitudes, with positive sign, $t_{ij} = \left|t_{ij}\right|$, which for atoms in optical lattices can be achieved either by coupling to an artificial magnetic field \cite{Jaksch2003, Sorensen2005,Lin2009b,Aidelsburger2011,Jimenez2012} or by lattice shaking \cite{Eckardt2010,Struck2011a,Sacha2011,Struck2012,Hauke2012c}. 
In the following, we will focus on the limit of infinite repulsion $U\to\infty$ and half filling $\langle n_i \rangle = 1/2$, under which the Bose--Hubbard model maps to the $S=1/2$ XY Hamiltonian
\begin{equation}
 \label{eq:HS}
  {\cal H_{\text{S}}}=
  \sum_{\braket{i,j}} t_{ij} \left(S_i^{\hspace{0.05cm}x} {\hspace{0.05cm}} S_j^{\hspace{0.05cm}x}
   + S_i^{\hspace{0.05cm}y} {\hspace{0.05cm}} S_j^{\hspace{0.05cm}y}\right)
\end{equation}
where $S_i^{\hspace{0.05cm}\alpha}$ are spin-$1/2$ operators acting on site $i$. 
\begin{figure}
	\centering
	\includegraphics[width=0.49\textwidth]{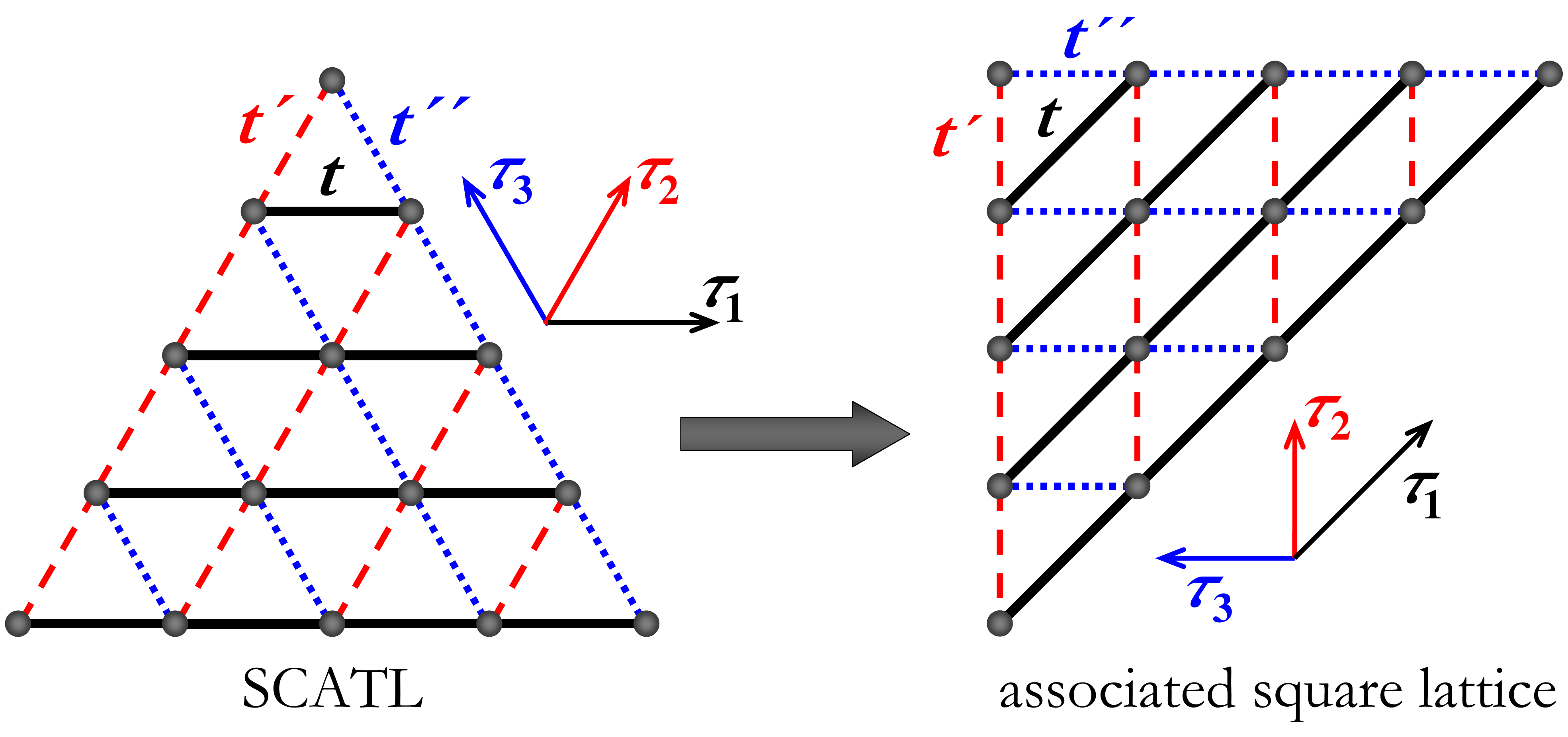}
	\caption{
	  {\bf Geometry of the SCATL.} The spins (gray bullets) are coupled to nearest-neighbors along the lattice vectors $\vs{\tau}_{1,2,3}$ by the couplings $t_{\vs{\tau}_{1}}\equiv t$, $t_{\vs{\tau}_{2}}\equiv t'$, and $t_{\vs{\tau}_{3}}\equiv t''$, which can all be mutually different. 
	  The right part of the Figure shows the associated square lattice.
	  The shown geometry is the one used in the ED of the 15-site system, chosen for maximal symmetry between all three couplings. 
	  \label{fig:geometry}
	}
\end{figure}

Recently, periodical driving of an optical lattice has allowed to experimentally realize a particular paradigmatic frustrated spin model, the spatially anisotropic triangular lattice (SATL) \cite{Struck2011a}. As described in Ref.~\cite{Eckardt2010}, the bare tunneling matrix elements $-\left|\tilde{t}_{ij}\right|$, connecting vertices of a triangular lattice such as depicted in Fig.~\ref{fig:geometry}, are dressed by the periodic driving with the factor $\left<\left<\ue^{\ri a_{ij}}\right>\right>$. Here, the double brackets denote time averaging and $a_{ij}=\bA\cdot\br_{ij}/\hbar$ is the projection of the periodic driving orbit, $\bA$, on the vector $\br_{ij}$ connecting sites $i$ and $j$. For brevity, we suppressed the explicit time dependence in  $a_{ij}$ and $\bA$.
In Ref.~\cite{Struck2011a}, this has been used to realize the AFM SATL by flipping the sign of the bare tunneling ($-\left|\tilde{t}_{ij}\right|\to\left|t_{ij}\right|$), thus creating AFM interactions leading to geometrical frustration. Further, the periodic driving orbit has been chosen such that the tunnelings along the diagonals are equal, i.e., $t'=t''$, with $t'$ along $\vs{\tau}_2=\left(1/2,\sqrt{3}/2\right)$ and $t''$ along $\vs{\tau}_3=\left(-1/2,\sqrt{3}/2\right)$, and different from the third, horizontal one [$t$, along $\vs{\tau}_1=\left(1,0\right)$]. 
The experiment of Ref.~\cite{Struck2011a} was at weak $U$, which corresponds to the classical limit, but expectations are high that soon also the regime of strong repulsion $U\to\infty$ may be reached. 
Here, the XY SATL is particularly interesting, because of predictions for several non-magnetic phases without classical counterpart \cite{Schmied2008, Hauke2010}.

Another implementation of the XY SATL within reach of current technology are trapped-ion setups \cite{Schmied2008}. Here, the bosonic particles governed by Eq.~\eqref{eq:HS} are provided by the vibrational modes of the ion crystal. The tunneling matrix elements $t_{ij}$ can be tuned via the preferred direction of vibration of the ions, and superposing an optical lattice allows to create strong on-site repulsion $U$. Proof-of-principle experiments using trapped ions to simulate spin models without \cite{Friedenauer2008,Islam2011} and also with frustration \cite{Kim2010} let it seem realistic that in the near future the SATL with a mesoscopic number of sites can be implemented. 

Motivated by these developments, we consider in this work a generalization of the SATL, the spatially completely anisotropic triangular lattice (SCATL), where all three tunneling matrix elements are different. 
Ultracold-atom experiments as Ref.~\cite{Struck2011a} and trapped-ions implementations after the proposal~\cite{Schmied2008} can easily be generalized to this situation by choosing an elliptical driving orbit or a preferred direction of ion vibration, respectively, which is not parallel to any of the sides of the triangular plaquettes. 
The geometry is depicted in Fig.~\ref{fig:geometry}, where we also sketch the associated square lattice with an interaction along one of the diagonals. We will work in the latter, to simplify the interpretation of our results.

The SCATL is interesting for two reasons. First, from a practical point of view, it is relevant for studying the sensitivity of the predicted non-magnetic phases to imperfect driving, i.e., driving which does not create two perfectly equal couplings. 
Second, from a more fundamental point of view, this model allows to investigate in a more general setting under which circumstances quantum-disordered phases appear. Namely, in the SATL, it was found that -- similarly to the same model with Heisenberg interactions -- non-magnetic ground states do not occur at largest frustration but instead intervene in the transition between phases with different type of order, which are therefore never directly connected. 
This leads also to the particularly important possibility that the spiral phase (occurring in the SATL near the isotropic point $t=t'=t''$) is completely surrounded by gapped quantum-disordered phases. This would mean that the two, supposedly different, gapped non-magnetic phases appearing in the SATL are actually continuously connected! Such a finding would yield deep insight into the nature of the quantum-disordered regions in the SATL. 

To address these issues, we investigate in this work the $S=1/2$ SCATL with AFM XY-interactions, Eq.~\eqref{eq:HS}, within Takahashi's modified spin-wave theory (MSWT) \cite{Takahashi1989}, supplemented with ordering-vector optimization. As shown previously \cite{Hauke2010,Hauke2011}, this improves significantly  over conventional spin-wave theory (as well as over conventional MSWT), as it allows to account for the dramatic quantum corrections to the type of order appearing in frustrated quantum antiferromagnets. Further, the breakdown of the theory provides a strong signal that the true ground state might be quantum disordered; hence, this method serves to efficiently find candidate models for spin-liquid behavior. 
While the main focus of this article is on the MSWT results, we complement them with exact diagonalization (ED) of small clusters.
The $S=1/2$ AFM SCATL with Heisenberg interactions, motivated by experiments on magnetic organic salts, has been treated in a similar way in the preceding article~\cite{Hauke2012a}.

The rest of this paper is organized as follows. 
First, to form intuition about the quantum SCATL, we discuss its classical counterpart (Sec.~\ref{cha:phd_classical}), and review important results from its well-studied limiting case, the SATL (Sec.~\ref{cha:KnownResults}). 
Sec.~\ref{cha:phd_quantum} contains our main results, namely the discussion of the quantum-mechanical ground-state phase diagram of the SCATL, including various observables from MSWT and ED, as well as, for a possible comparison to experiment, the expected boson momentum distributions at selected points of the phase diagram. 
We delegate the technical details of the MSWT to the Appendix.
Sec.~\ref{cha:conclusion}, finally, provides some conclusions.

\subsection{Classical phase diagram\label{cha:phd_classical}}

In this section, we discuss the classical phase diagram of the SCATL, which can serve as a guide to what ordered phases are to be expected, and which allows to appreciate the changes brought about by quantum fluctuations. 

To obtain the classical solution, we replace the quantum-XY spins in Eq.~\eqref{eq:HS} by classical rotors in the $xy$-plane. The ordering vector $\vect{Q}^{\mathrm{cl}}=\left(Q_{x}^{\mathrm{cl}},Q_{y}^{\mathrm{cl}}\right)$ is found as the $\vect{k}$-vector which minimizes the Fourier transform of the coupling strengths. It fixes the direction of each spin (up to a global phase) as 
$\vect{S}_{\vect{i}}=S\left(\cos(\vect{Q}^{\mathrm{cl}}\cdot\vect{r}_i),\sin(\vect{Q}^{\mathrm{cl}}\cdot\vect{r}_i)\right)$. 
We find
\eqa{
  Q_{x}^{\mathrm{cl}}& = &\left\{\begin{array}{l}
                              \pi \quad \mathrm{for} \quad -\frac{t}{2t'}-\frac{t'}{2t}+\frac{t t'}{2t''^2} \leq 1 \\
			      0   \quad \mathrm{for} \quad -\frac{t}{2t'}-\frac{t'}{2t}+\frac{t t'}{2t''^2} \geq 1 \\
			      \arccos\left( -\frac{t}{2t'}-\frac{t'}{2t}+\frac{t t'}{2t''^2} \right) \quad \mathrm{else}
                             \end{array}
			\right. \\
  Q_{y}^{\mathrm{cl}}& = &\left\{\begin{array}{l}
                              \pi \quad \mathrm{for} \quad -\frac{t}{2t''}-\frac{t''}{2t}+\frac{t t''}{2t'^2} \leq 1 \\
			      0   \quad \mathrm{for} \quad -\frac{t}{2t''}-\frac{t''}{2t}+\frac{t t''}{2t'^2} \geq 1 \\
			      \arccos\left( -\frac{t}{2t''}-\frac{t''}{2t}+\frac{t t''}{2t'^2} \right) \quad \mathrm{else}
                             \end{array}
			\right.
			      \nonumber
}

\begin{figure}
	\centering
	\includegraphics[width=0.49\textwidth]{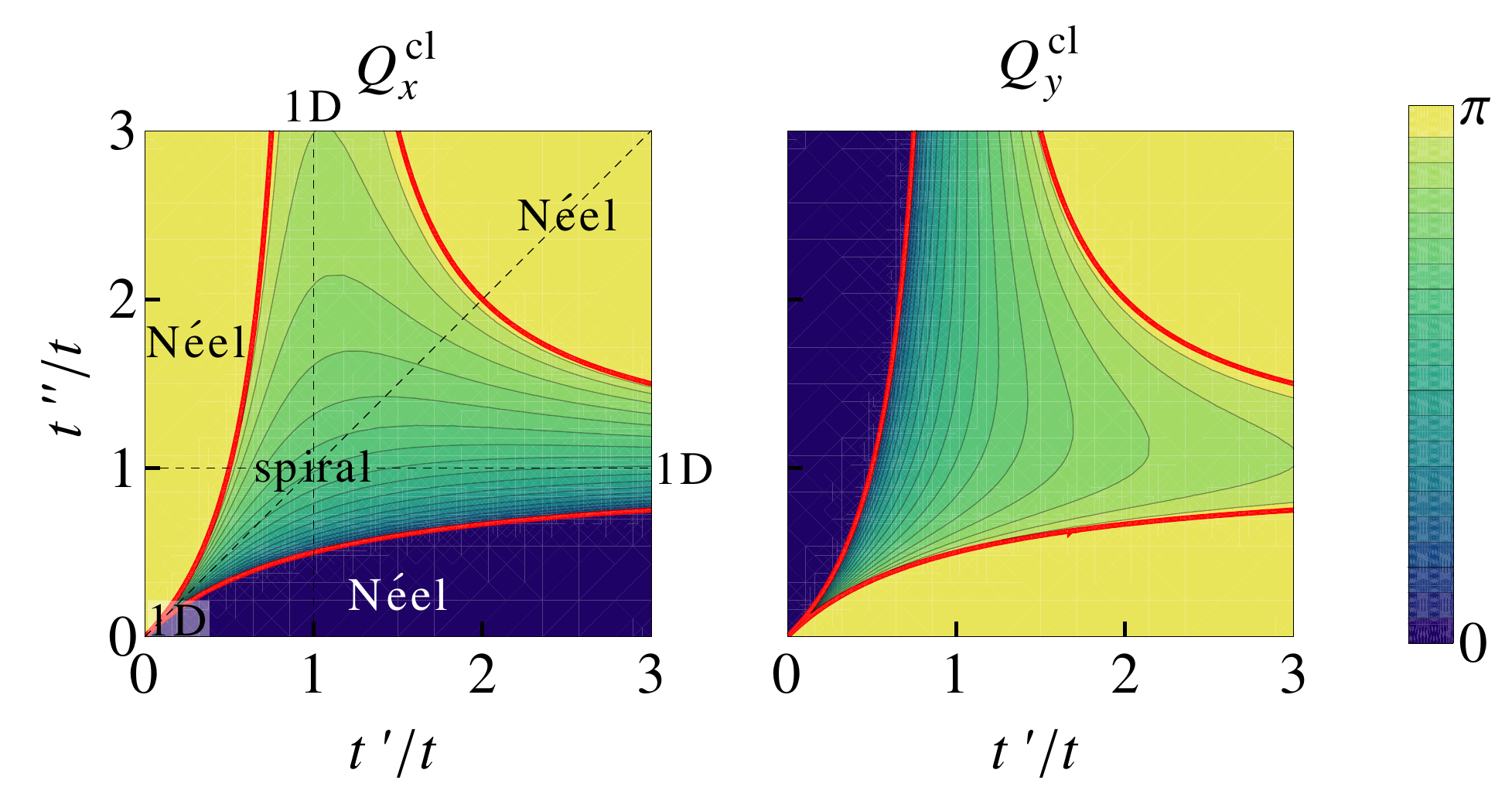}
	\caption{
	  {\bf Classical phase diagram of the SCATL: The ordering vector} evidences three N\'eel-ordered phases, an extended spiral-ordered phase, and limits where the system decouples into an ensemble of independent chains, as indicated by the labels in the left panel.
	  The thick red lines denote transitions between different kinds of order, and along the dashed black lines the system is in SATL limit.
	  \label{fig:phd_classical}
	}
\end{figure}

The classical phase diagram of the SCATL, plotted in Fig.~\ref{fig:phd_classical}, contains several N\'eel-ordered phases and an extended spiral-ordered phase. 
The N\'eel phases spread around the square-lattice limits [$(t'/t,t''/t)=(1, 0)$ with $\vect{Q}^{\mathrm{cl}}=\left(0,\pi\right)$, [$(t'/t,t''/t)=(0, 1)$ with $\vect{Q}^{\mathrm{cl}}=\left(\pi,0\right)$, and $t'/t,t''/t\gg1$ with $\vect{Q}^{\mathrm{cl}}=\left(\pi,\pi\right)$]. 
The spiral phase, with continuously varying ordering vector, connects smoothly to the N\'eel phases, and occupies the extended region between them. In particular, it extends all the way to $t'/t=t''/t=0$ [and, symmetrically, to ($t'/t=1$, $t''/t\to\infty$) and ($t''/t=1$, $t'/t\to\infty$)], where the system decouples into an ensemble of 1D chains. 

\subsection{Known results in limiting cases\label{cha:KnownResults}}

In this section, we discuss well-known limiting cases of the quantum SCATL, including results on the SATL. This helps us to assess which phases and quantum effects to expected in the phase diagram of the SCATL.

The SCATL interpolates between several important limiting cases. For $(t'/t,t''/t)=(1, 0)$, $(t'/t,t''/t)=(0, 1)$, and $t'/t,t''/t\gg1$, one recovers the square lattice limit. Here, N\'eel long-range order (LRO) persists in the quantum case \cite{Sandvik1999}. Similarly, in the isotropic triangular lattice, $t'=t''=t$, spiral LRO survives quantum fluctuations \cite{Momoi1994}. 
The limits ($t'=t''=0$), ($t'\to\infty$ with $t''=\mathrm{const}$), and ($t''\to\infty$ with $t'=\mathrm{const}$) correspond to ensembles of decoupled, critical XY-chains, with algebraic correlations along individual chains but no correlations between them.

\begin{figure}
	\centering
	\includegraphics[width=0.49\textwidth]{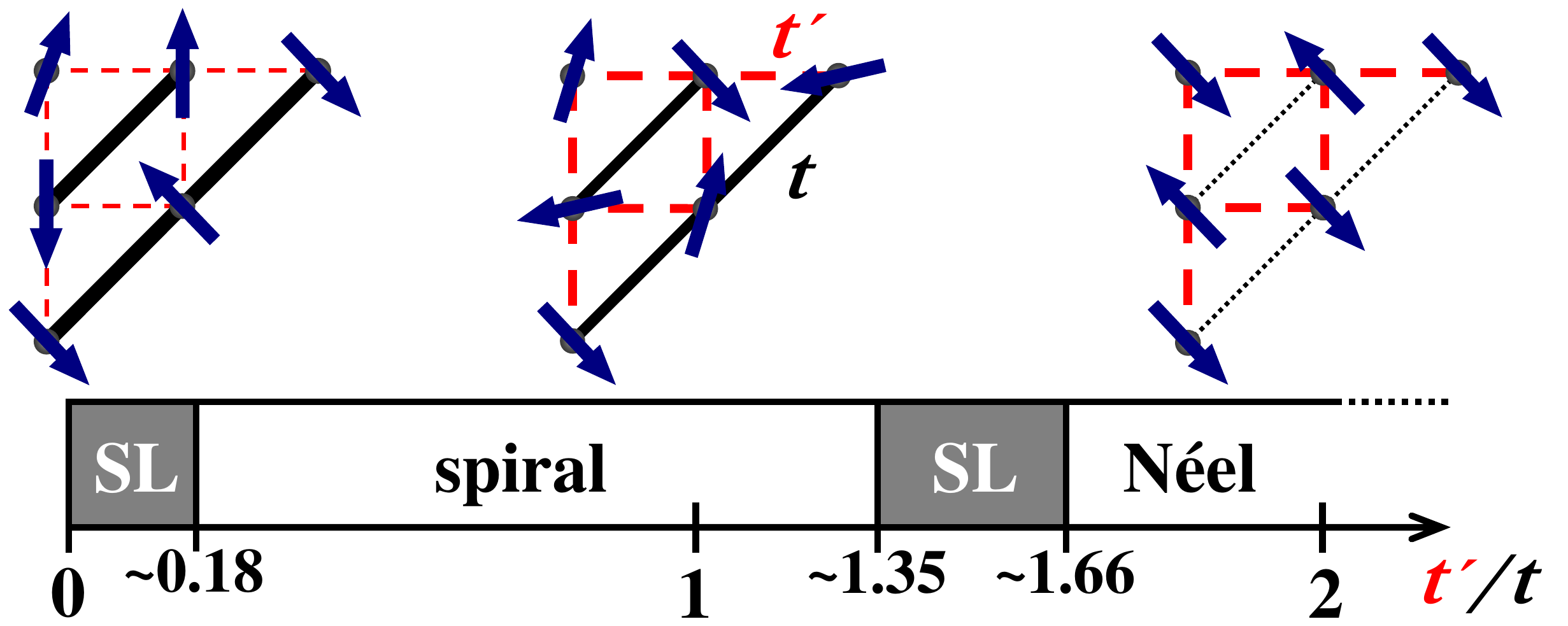}
	\caption{
		{\bf The MSWT quantum phase diagram of the SATL} (from Ref.~\cite{Hauke2010}) contains N{\'e}el order (which is considerably more stable than in the classical model), spiral order (which is destabilized by quantum fluctuations), and two putative spin-liquid (SL) phases. These are found through the breakdown of the theory and a disappearing  spin stiffness, which indicates a gapped disordered phase. In the purely 1D limit, MSWT recovers convergence and produces the 1D critical state. 
		We include sketches of classical states, where blue arrows indicate the directions of the classical rotors, namely, the 1D state at $t'/t=0$, the spiral state at $t'/t=1$, and the 2D-N\'{e}el state at $\alpha\geq 2$, . 
	  \label{fig:phd_SATL}
	}
\end{figure}

For $t'=t''\equiv\alpha t$ (or, equivalently, $t'=t$ or $t''=t$), one recovers the SATL, where projected entangled-pair states (PEPS) calculations \cite{Schmied2008} and MSWT (as well as ED) \cite{Hauke2010} suggest the existence of several non-magnetic phases.
In the rest of this section, we briefly review the main features of the SATL phase diagram, proceeding from large to small $\alpha=t'/t$ (for comparison, Fig.~\ref{fig:phd_SATL} reproduces the MSWT phase diagram found in Ref.~\cite{Hauke2010}). The qualitative behavior is very similar to the SATL with Heisenberg interactions (for a short review, see preceding publication~\cite{Hauke2012a}), but the decreased symmetry in the XY case mitigates the effects of quantum fluctuation.

At large $\alpha$, order-by-disorder effects due to quantum fluctuations stabilize N{\'e}el order considerably, moving the point where it disappears downwards from the classical value $\alpha=2$ to values between $\alpha\approx1.4$ (PEPS, Ref.~\cite{Schmied2008}) and $\alpha\approx1.66$ (MSWT, Ref.~\cite{Hauke2010}).
These methods predict that quantum fluctuations spread the transition point between the N{\'e}el and the spiral phase into a quantum-disordered phase. In the following, we term this predicted disordered region ``large-$\alpha$ quantum disorered region'' (large-$\alpha$ QDR). 
One of the main aims of this paper is to investigate if the fact that a quantum-disordered phase intervenes in transitions between commensurate and incommensurate order is a more general feature of frustrated quantum antiferromagnets. 

Similar behavior has been found in a variety of quantum spin models, including $J_1J_2J_3$-models on the square lattice \cite{
Chandra1988,Locher1990,Ferrer1993,Zhong1993,Leung1996,Capriotti2004a,Capriotti2004b,Mambrini2006,Shannon2006,
Murg2009,Richter2010,Hauke2011,Reuther2011b},  
and frustrated honeycomb models with Heisenberg
\cite{Farnell2011,Albuquerque2011,Li2012}
or XY interactions
\cite{Varney2011}. 
In fact, since quantum phase transitions are driven by quantum fluctuations, one might expect that -- if anywhere -- a complete restructuring of the ground state in favour of a quantum mechanical configuration may occur preferably close to a quantum critical point. 
It, hence, seems plausible that at such points quantum fluctuations are most effective in disrupting classical order. 
Indeed, a similar effect occurs in classical statistical physics. Assume that there is a transition between a commensurate and an incommensurate phase which both show LRO. As a first notable thing, close to this transition the thermal phase transition to a disordered state will typically happen at lower temperature than far away from it. Beyond the thermal phase transition, the disordered phase will show short-range order of the type corresponding to the adjacent long-range ordered phase. The transition between the two different kinds of short-range order is called a \emph{disorder point} \cite{Stephenson1969,Stephenson1970,Stephenson1970a,Stephenson1970b}. 
Interestingly, the correlation length associated with the two kinds of short-range order has a minimum just at this point. 
Hence -- similar to what is found in the quantum models -- thermal fluctuations tend to suppress order most effectively at a commensurate--incommensurate transition. 

At the low-$\alpha$ side of the spiral phase, Ref.~\cite{Schmied2008} predicts a gapped disordered phase for as large $\alpha$ as $\approx 0.6$, followed by a gapless disordered phase in the region $0\leq\alpha\lesssim 0.4$. 
In the following, we term this predicted disordered region ``small-$\alpha$ QDR.'' 
Similarly, MSWT computations indicate a quantum disordered phase \cite{Hauke2010} below $\alpha\approx0.18$, but the very small spin stiffness at already $\alpha\approx0.35$ suggests that taking quantum fluctuations into account more completely than within MSWT could destabilize the weak magnetic order already at larger $\alpha$. 

\section{Quantum-mechanical phase diagram\label{cha:phd_quantum}}

From the discussion of the classical phase diagram and the limiting cases, we have the necessary background to tackle the quantum-mechanical ground-state phase diagram of the XY SCATL. 
We compute it using MSWT supplemented with ordering-vector optimization, working directly in the thermodynamic limit.
Since this method is described in detail in our previous papers \cite{Hauke2010,Hauke2011}, we delegate the technical aspects to the Appendix, and only summarize here the main idea. The starting point is a classical state, which one dresses with quantum fluctuations in a second-order spin-wave expansion. This yields a bosonic Hamiltonian, the ground state of which is found self-consistently by minimizing its mean-field free energy. For this, quartic terms, i.e., interactions between spin waves, are decoupled via Wick's theorem. 
Additionally, we employ Takahashi's modification of vanishing magnetization. This constricts the average number of spin-wave excitations to a physical value, in contrast to conventional spin-wave theory, where the spin-wave excitations grow completely unchecked. 
This modification has proven a crucial improvement to describe low-dimensional systems with weak order tendencies. 

Typically, one uses the classical ground state as the reference state. However, in many models quantum fluctuations considerably shift the type of predominant order. Therefore, we find the ordering vector giving the best classical reference state by including it in the self-consistent optimization. This has proven crucial to capture, e.g., the stabilization of the N\'eel phase by quantum fluctuations. 
As has been proposed in Refs.~\cite{Hauke2010,Hauke2011}, the breakdown of the theory strongly suggests that at mean-field level \emph{no} semi-classical reference state yields a good description of the quantum ground state. This is then interpreted as an indication of quantum-disordered behavior in the true ground state. This will be an important aspect for the interpretation of the quantum phase diagram. 

We compare these results to exact diagonalization (ED) of a 15-site lattice, as depicted in Fig.~\ref{fig:geometry}. 
The geometry is chosen for its symmetry between $t$, $t'$, $t''$ bonds. 
It is important to leave the boundaries open to allow for incommensurate ordering vectors.

\subsection{MSWT and ED results -- ordering vector and order parameter \label{cha:quantumPhaseDiagramResultsEDandMSWT}}

In this section, we give a first overview over the phase diagram, as obtained from the ordering vector $\vect{Q}$ and the order parameter $M$, followed in the next two sections by more detailed analyses. 
In MSWT, ordering vector and order parameter are direct results of the optimization [see Appendix, Eqs.~\eqref{Qs} and~\eqref{eq:orderParameterMSWT}]. In ED, they can be extracted from the static structure factor, 
\eq{
\label{eq:structureFactor}
S(\vect{k})=\frac{1}{N^2}\sum_{i,j} \ue^{i \vect{k}\cdot(\vect{r}_i-\vect{r}_j)} \braket{S^x_i S^x_j+S^y_i S^y_j}\,.
}
Its peak lies at the ordering vector $\vect{Q}^{\mathrm{ED}}$, and the square root of its height, $\sqrt{S(\vect{Q}^{\mathrm{ED}})}\equiv M^{\mathrm{ED}}$, approaches in the thermodynamic limit the order parameter $M$. 

\begin{figure}
	\centering
	\hspace*{-0.5cm}\includegraphics[width=0.52\textwidth]{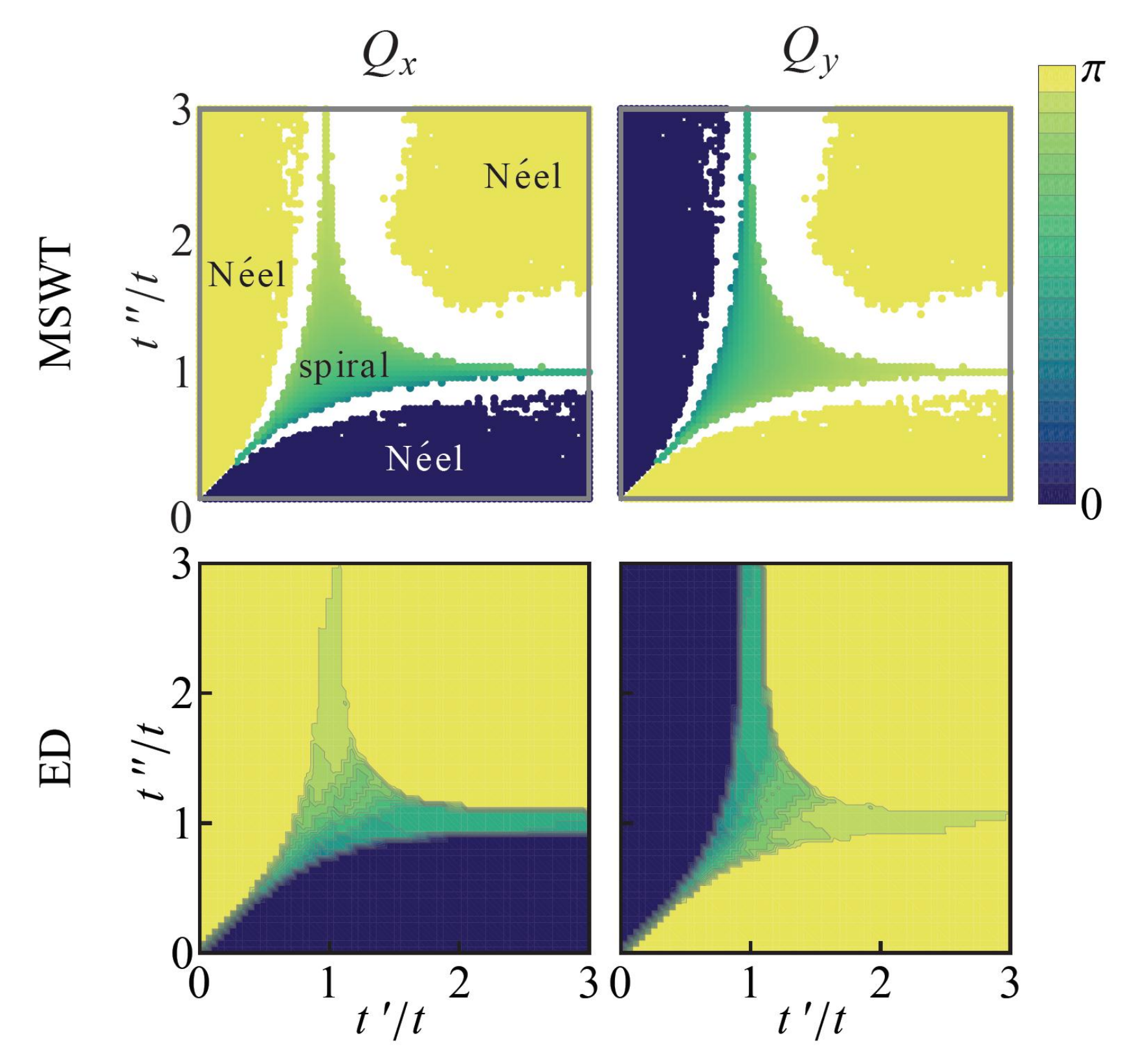}
	\caption{
	  {\bf Quantum-mechanical phase diagram of the SCATL, ordering vector.} 
	  {\bf Upper row: MSWT data.} Quantum fluctuations stabilize the N\'eel phase. Around $t'\approx t'' \approx t$, a part of the classical spiral phase survives quantum fluctuations (labels in the upper left panel). 
	  {\bf Lower row: ED data} for $N=15$ sites. Already for this small system, it can be appreciated that the N\'eel phase grows at the expense of spiral order (compared to the classical case) .
	  \label{fig:phd_quantum_Q}
	}
\end{figure}
\begin{figure}
	\centering
	\hspace*{-1cm}\includegraphics[width=0.55\textwidth]{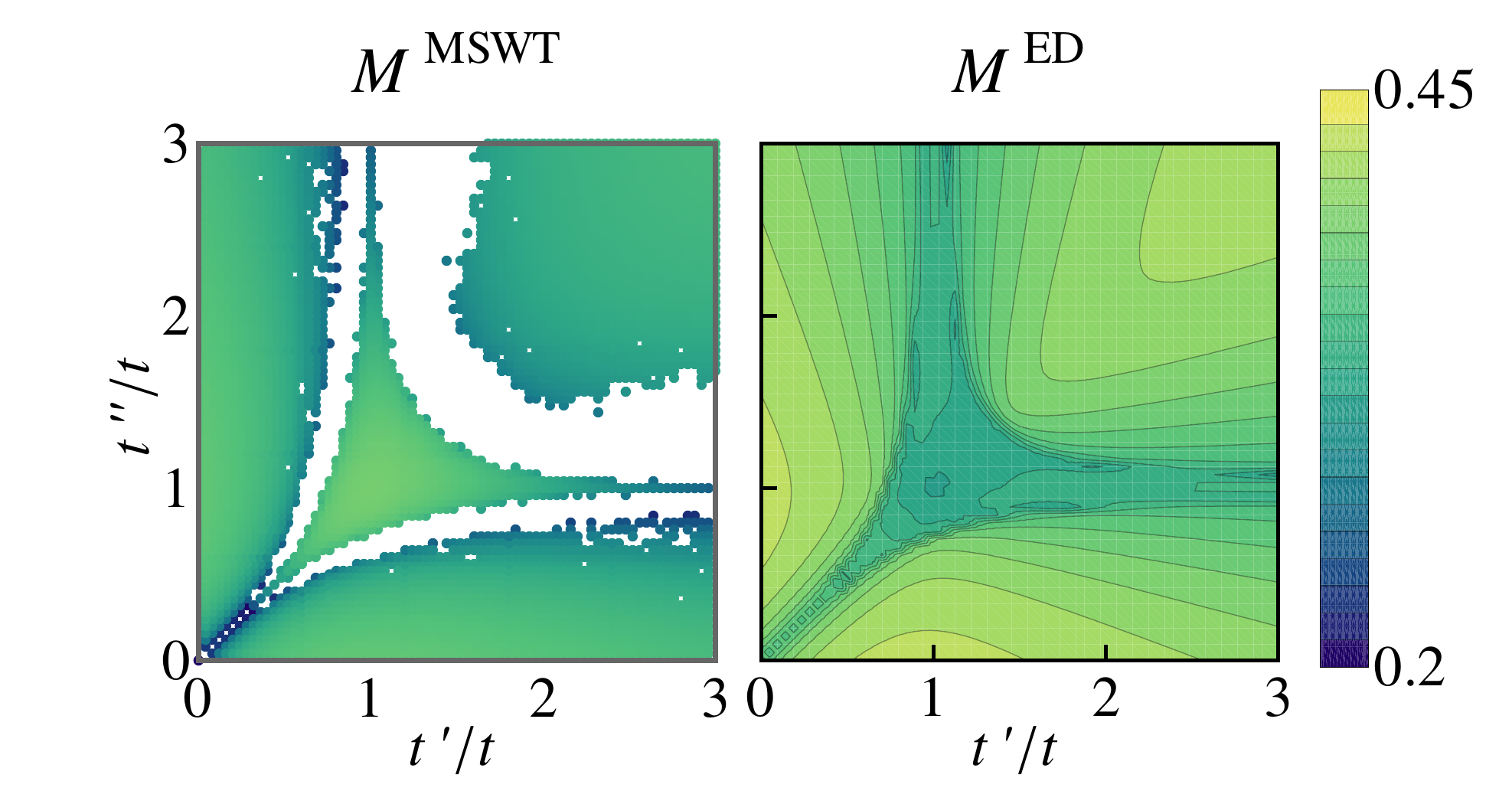}
	\caption{
	  {\bf Quantum-mechanical phase diagram, order parameter.} ED results qualitatively confirm MSWT. In particular, the order parameter for both methods decreases rapidly upon approaching the MSWT breakdown regions. 
	  \label{fig:phd_quantum_M}
	}
\end{figure}
As seen in the MSWT and ED ordering vectors, presented in Fig.~\ref{fig:phd_quantum_Q}, quantum fluctuations stabilize the N\'eel phases compared to the classical case, as already observed in the SATL (Sec.~\ref{cha:KnownResults}).
In the central region around $t'\approx t''\approx t$, a broad range of incommensurate ordering vectors indicates spiral order. 
The finite MSWT order parameter (Fig.~\ref{fig:phd_quantum_M}, left panel) shows that in these phases indeed LRO survives quantum fluctuations.
(Note that the self-consistent MSWT calculations become relatively unstable for small order parameters, which results in ragged phase boundaries.)

In the N\'eel phases, the ED order parameter (Fig.~\ref{fig:phd_quantum_M}, right panel) is maximal, giving support to the assumption that here LRO persists. However, it is much smaller in the spiral phase than the MSWT value, a discrepancy already found in the SATL~\cite{Hauke2010}. (This could be due to corrections to the third-order spin-wave expansion which our approach neglects, and which can become important in spiral configurations  \cite{Chernyshev2009}.)

Between the ordered phases, we find a broad region where MSWT breaks down, indicating as usual \cite{Hauke2010,Hauke2011} that these regions do not allow a description in terms of an ordered, semi-classical state. 
Therefore, it appears that it is a quite universal feature of frustrated quantum antiferromagnets that spiral- and collinearly-ordered phases are always separated by quantum disordered phases. This finding constitutes the main result of this paper. 

The strong decrease of the MSWT and ED order parameters (Fig.~\ref{fig:phd_quantum_M}) upon approaching this region gives support to this interpretation (which we will further corroborate in the next two sections). 
Note also that both the ED and MSWT order parameter seem to disappear relatively smoothly when approaching the putative 1D-like QDR (consider, e.g., in the range $2\lesssim t'/t\lesssim 3$, $t''/t\to 1^-$). Upon approaching the putative large-$\alpha$ QDR dividing spiral from N\'eel LRO, on the other hand, for ED, the order parameter decreases more sharply (consider, e.g., the line $t'/t=1$, $t''/t\to 1^-$).  Here, for MSWT, the breakdown occurs abruptly at finite order parameters. 
This points at a difference in the type of phase transition upon approaching the large-$\alpha$ QDR and the non-magnetic region at the decoupled-chains limit.

The rest of this article is devoted to fleshing out our main finding, the appearance of a disordered region encircling the spiral phase.

\subsection{Supporting observables from MSWT -- spin stiffness and spin-wave velocities \label{cha:spinStiffnessAndSpinWaveVelocities}}

\begin{figure}
	\centering
	\hspace*{-3cm}\includegraphics[width=0.35\textwidth]{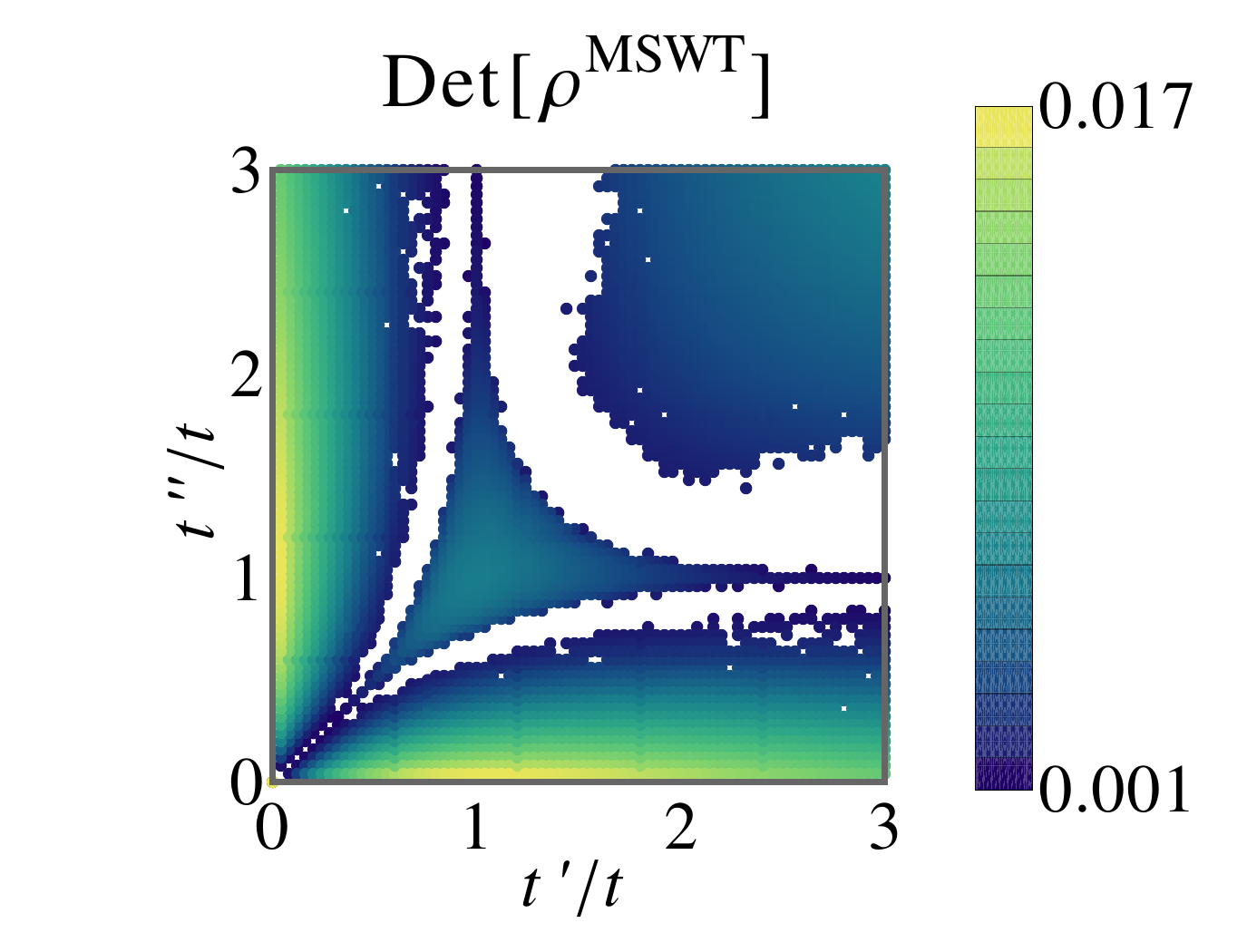}\\
	\includegraphics[width=0.49\textwidth]{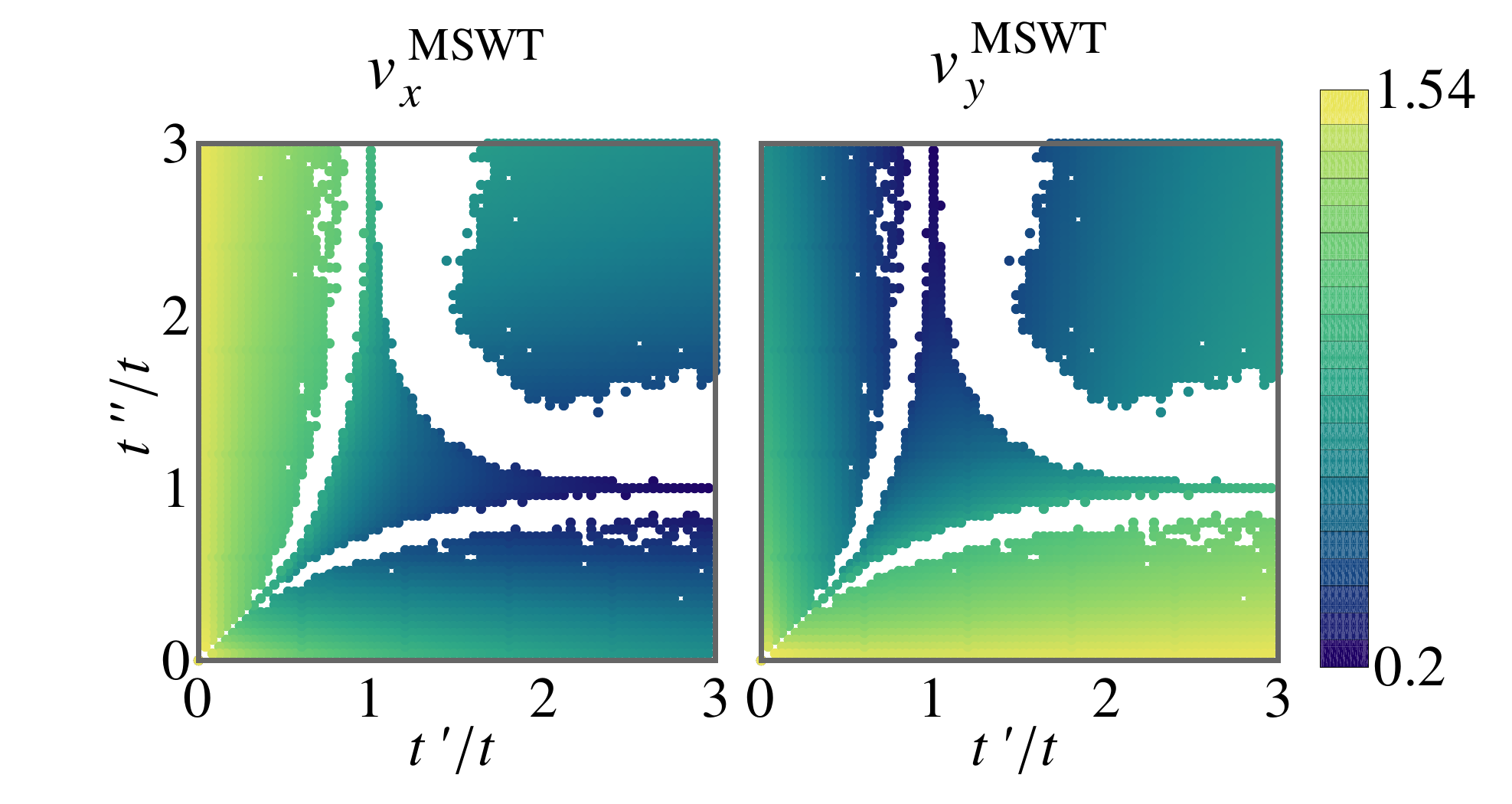}
	\caption{
	  {\bf Upper panel: The partial spin stiffness} decreases upon approaching the MSWT breakdown region, suggesting the disruption of magnetic LRO.
	  {\bf Lower panels: The spin-wave velocities} perpendicular to the dominating coupling strength soften in the 1D limits.
	  Differences in the spin-wave velocities might allow to measure the anisotropy of the SCATL.
	  All quantities are normalized to $1+t'+t''$. 
	  \label{fig:phd_MSWT_spinvelocityAndStiffness}
	}
\end{figure}
In Refs.~\cite{Hauke2010,Hauke2011}, the spin-stiffness tensor, which characterizes the stiffness of the magnetic order under change of the ordering vector, has proven a valuable consistency check of our MSWT calculations. 
Its components are 
\eq{
  \label{eq:spinstiffness}
  \rho_{\alpha,\beta}=\frac{d^2 \mathcal{F}}{d Q_\alpha d Q_\beta}\,,
}
where $\mathcal{F}$ is the free energy. Even if the order parameter is finite, a small spin stiffness suggests that further quantum fluctuations than taken into account within MSWT could disrupt the remaining order
\footnote{Also, if the spin stiffness is small when approaching the MSWT breakdown region, we are led to assume that the breakdown is not caused by numerical problems, but that it is a physical effect, i.e., due to the lack of a description in terms of a semi-classically ordered reference state.}.

Since for our purposes an upper bound for the spin stiffness is sufficient, we take the partial derivative in Eq.~\eqref{eq:spinstiffness}. The exact spin stiffness can be computed via the total derivative. To this, within the self-consistent MSWT calculations, one first has to find the optimal ordering vector. Then, one reruns the self-consistent MSWT equations for several fixed, slightly non-optimal ordering vectors, yielding slightly larger energies. The spin stiffness can be derived by fitting a quadratic form to the resulting energy landscape. In the self-consistent iteration, the mean fields characterizing the MSWT state can adjust to a changed ordering vector. This effect is not taken into account in the partial derivative, which hence provides an upper bound to the total spin stiffness. We find that it suffices to extract the location of disordered phases, but it may yield wrong results about their nature. 
In particular, we found in the SATL \cite{Hauke2010} that, upon approaching the putative small-$\alpha$ QDR, not only the total inter-chain, but also the total intra-chain spin stiffness decreases strongly. Since such a behavior is not consistent with algebraic correlations along the chains, this can be interpreted as an indication of a gapped QDR. The partial spin stiffness computed in Ref.~\cite{Hauke2010}, on the other hand, only vanishes in the inter-chain direction. Hence, it may not be able to distinguish gapped from gapless spin liquids. However, it still seems to adequately capture the location of disordered regions.

In Fig.~\ref{fig:phd_MSWT_spinvelocityAndStiffness}, upper panel, we show the determinant of the spin-stiffness tensor, $\det(\rho)$, normalized to the coupling strengths $1+t'+t''$.
As we should expect \cite{Chubukov1994}, it decreases upon approaching the phase transitions, especially from the N\'eel-ordered side.
At large $t'$ ($t''$), this decrease is due to a softening of the stiffness in $x$ ($y$) direction, and at small $(t'/t,t''/t)$ in the direction perpendicular to $\vect{\tau}_1$ (as has also been found in Ref.~\cite{Hauke2010}).

Another indicator for approaching disordered phases is given by the spin-wave velocities $v_{x,y}$, which can be connected to the spin stiffness via the susceptibility~\cite{Halperin1969}.
Since the spin-wave velocities are defined as the leading order of an expansion of the spin-wave dispersion relation, Eq.~\eqref{disp}, around small $\left|\vect{k}\right|$, i.e., 
\begin{subequations}
\eqa{
    v_x&=&\left.\lim_{k_x\to 0} \omega_{\vect{k}}/k_x\right|_{k_y=0} \,, \\
    v_y&=&\left.\lim_{k_y\to 0} \omega_{\vect{k}}/k_y\right|_{k_x=0} \,,
}
\end{subequations}
they can be measured directly from the spin-wave dispersion, allowing an experimental check of our findings. 

As seen in Fig.~\ref{fig:phd_MSWT_spinvelocityAndStiffness}, lower panels, close to the 1D breakdown region, they, too, soften in the direction perpendicular to the dominating coupling. 
On the other hand, when approaching the putative large-$\alpha$ QDR dividing the spiral from the N\'eel phase, both spin-wave velocities remain finite. This suggests that the large-$\alpha$ QDR could be qualitatively different from the non-magnetic phase found in the limit of decoupled chains.

\subsection{Supporting observables from ED -- energy derivative, gap, and chiral correlations}

\begin{figure}
	\centering
	\hspace*{-0.5cm}\includegraphics[width=0.55\textwidth]{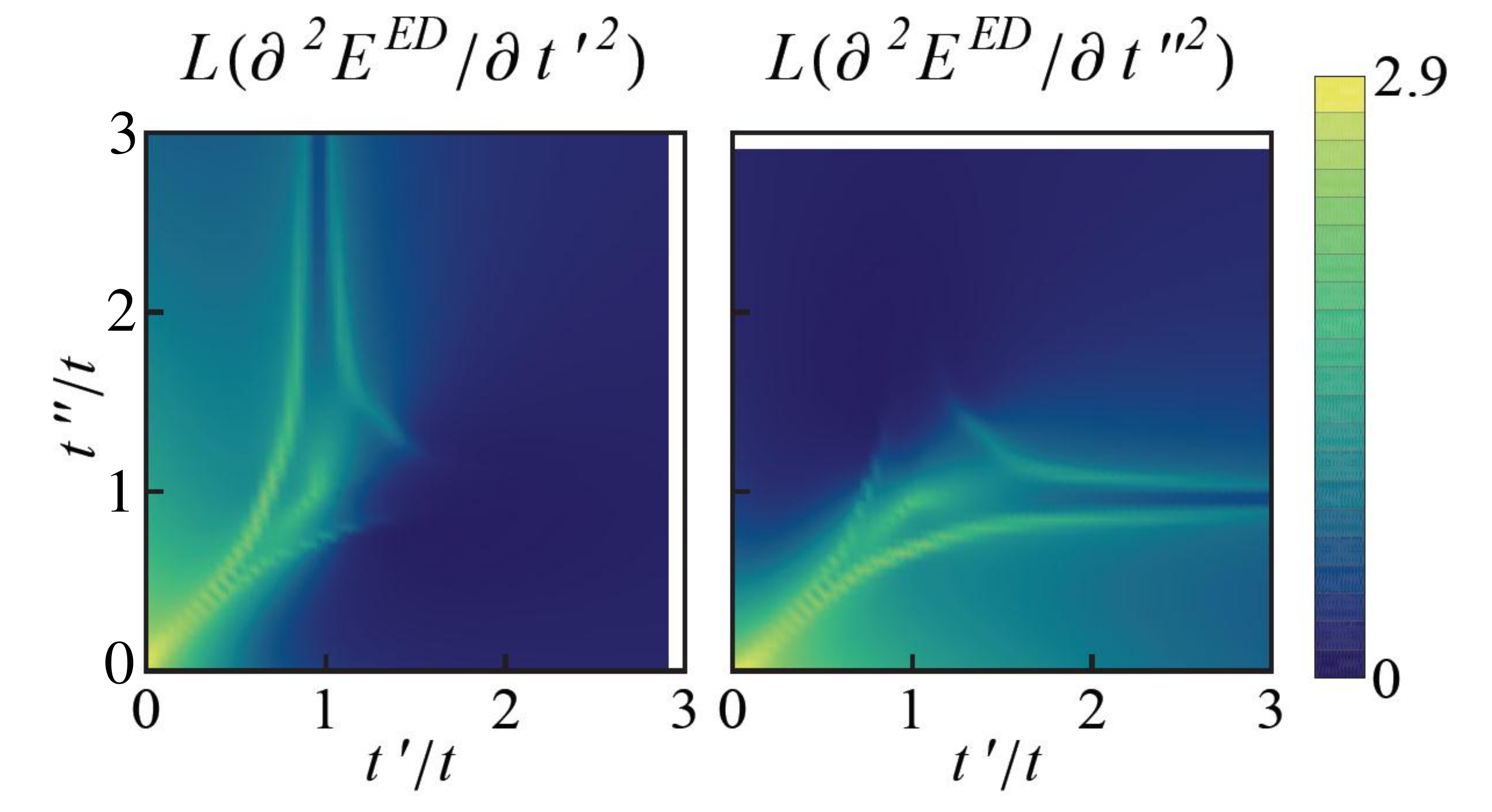}
	\caption{
	  {\bf Second derivative of ED ground-state energy} per spin for the $N=15$ system. For clarity, we plot the logarithm after a shift to values larger one, $L(\partial^2 E^{\mathrm ED} / \partial {t^\gamma}^{2})$, where $L(x)=\log(1+\max(x)-x)$ and where $t^{\gamma}$ is $t'$ or $t''$. Strong peaks clearly mark the phase transitions from the N\'eel phases. An additional peak around $(t',t'')=(1,1)$ might be an indication of an additional phase, separating the N\'eel phases from the spiral one. 
	  \label{fig:phd_ED_energy2ndDerivative}
	}
\end{figure}

The ED observables investigated in Sec.~\ref{cha:quantumPhaseDiagramResultsEDandMSWT} allowed to interpret the predominant ordering behavior, but did not yield clear evidence if within ED really quantum phase transitions exist, and if yes, where. 
The second derivative of the ED ground-state energy, which we plot in Fig.~\ref{fig:phd_ED_energy2ndDerivative}, can provide such an indicator, as in the thermodynamic limit it diverges at a quantum phase transition. 

Indeed, there are clear peaks along lines similar to where in MSWT the N\'eel order breaks down. Also, a peak appears around $(t',t'')=(1,1)$. 
This might be interpreted as the precursor of a quantum phase transition away from the spiral state, and possibly to the non-magnetic phase which is supposed to exist in this system. 

\begin{figure}
		\centering
		\includegraphics[width=\columnwidth]{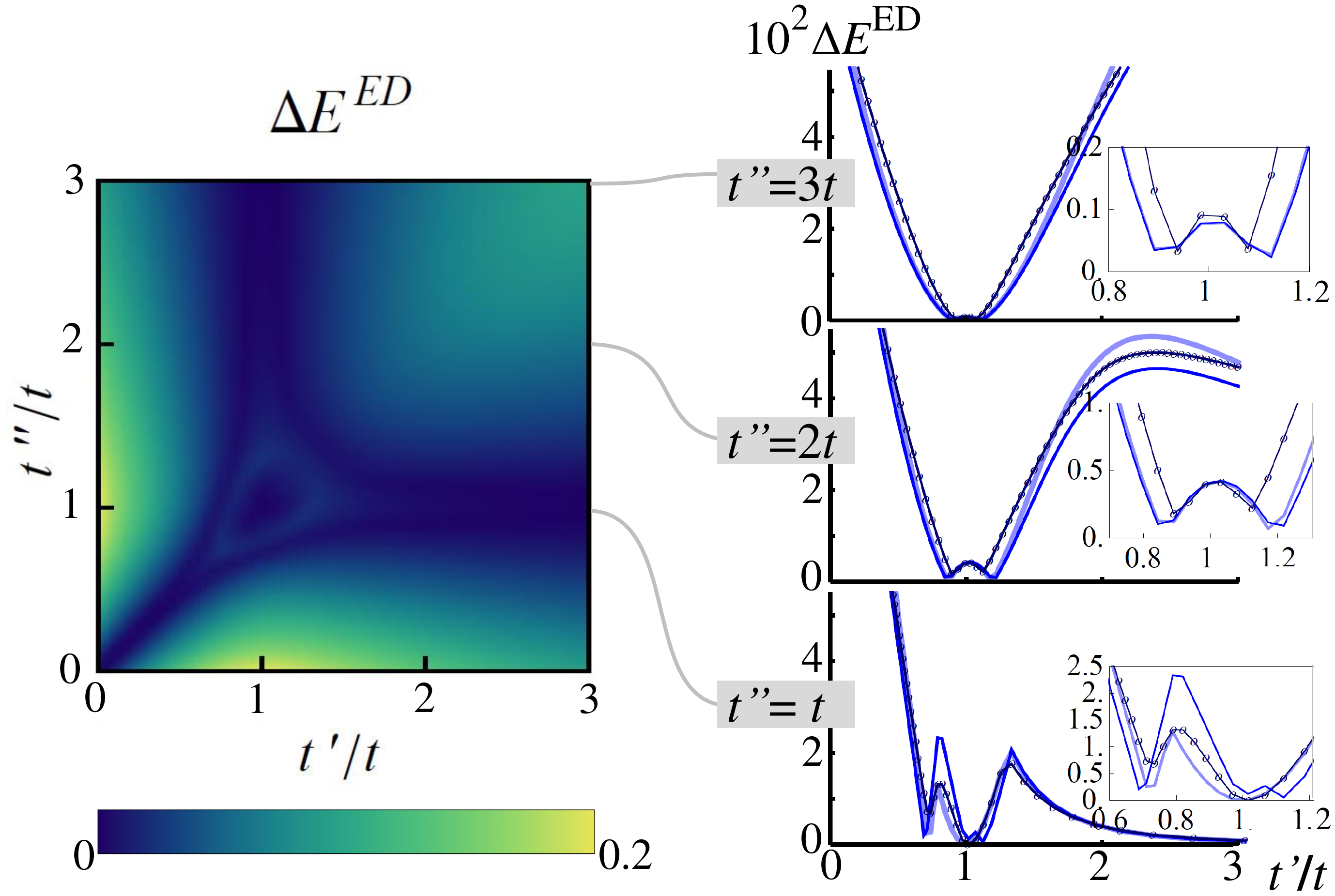}
	\caption{
	  {\bf Left panel: The singlet gap from ED} gives support to the MSWT phase diagram: A finite gap separates in the N\'eel phases spin-wave excitations from the ground state. In the spiral phase, the ground state is doubly degenerate due to the ambiguity in choice of chirality. The finite gap surrounding the degenerate region could be a precursor of a gapped, disordered phase. At the quantum phase transitions to the N\'eel phases, the gap closes again. 
	{\bf Right panels: cuts} at fixed $t''/t=1,2,3$ for triangles with increasing $N$ (from light to dark and thick to thin: 6,10,15). There is little size dependence in the central gapped phase ($t''=2,3t$ with $t'\approx t$, as well as $t''=t$ and $t'/t\gtrsim1.5$). 
	  \label{fig:phd_ED_gap}
	}
\end{figure}

We get further support for this phase diagram from the ED energy gap per spin between ground and first excited state, Fig.~\ref{fig:phd_ED_gap}. 
In the well-known limiting cases of the SCATL, it behaves as expected: 
There is no singlet gap close to the decoupled-chains limits, since the system is then in a critical phase. 
In the N\'eel ordered phases, there is a large gap which separates the ground state from closely-spaced excitations, which in larger lattices become the spin waves, collapsing slowly towards the ground state \cite{Lhuillier2005}. 
This is consistent with the considerable size dependence found in our calculations for that parameter region, as can be seen in the right panels of Fig.~\ref{fig:phd_ED_gap}, where we plot cuts of $\Delta E^{\mathrm{ED}}$ at fixed $t''/t=1,2,3$ for triangular systems similar to the one in Fig.~\ref{fig:geometry} with $N=6,10,15$.

On the contrary, there is no gap in the spiral phase, 
because there are two degenerate ground states with opposite chirality 
\footnote{In the spiral phase, there is a gap, similar to the spin-wave gap of the N\'eel phases, between the \emph{second} and the \emph{third} energy level.}. 
We find that the vanishing of the gap depends strongly on the system geometry, but it occurs consistently for all triangular systems considered.

Interestingly, the gapless spiral phase is surrounded by a region where the gap attains considerable values. The very small dependence on system size for this parameter region indicates that this is stable towards the thermodynamic limit. A finite gap is not consistent with a spiral-ordered phase. On the other hand, the predominant order in this region is at incommensurate wave-vectors. Hence, the finite gap is clearly not  due to square-lattice N\'eel physics. 
Optimistically, these findings could therefore be interpreted as the precursors of a gapped non-magnetic phase phase. 
This gapped region completely encircles the spiral phase, suggesting that the low- and large-$\alpha$ gapped QDRs found in the SATL could actually be continuously connected via the additional anisotropy of the SCATL. 

The gap closes again upon approaching the N\'eel phases, indicating a quantum phase transition. 

To understand better the nature of this possible gapped, non-magnetic region, we now study the persistence of chiral correlations.
The vector chirality is defined as 
\eq{
  \kappa_{i,j,k}=\frac{2}{3\sqrt{3}}\left(\vect{S}_i\times\vect{S}_j+\vect{S}_j\times\vect{S}_k+\vect{S}_k\times\vect{S}_i\right)_z\,,
}
where the sites $\left\{i,j,k\right\}$ are located counter-clockwise on a triangle. For the small systems used in our ED, we generalize the chiral correlations \cite{Richter1991} to
\eq{
  \Psi_{-}=\frac{4}{N_{\Delta}}\braket{\sum_c s_c \kappa_c \sum_a s_a \kappa_a}\,.
}
Here, the sum $a$ runs over all triangles, while $c$ runs only over the central ones to reduce boundary effects. The factors $s_{a,c}$ weight $\kappa_{a,c}$ with a $+$ ($-$) sign if the triangle points upwards (downwards). 
The prefactor, where $N_{\Delta}$ is the number of summands, is chosen such that the chiral correlation has the same theoretical maximum of $\frac{9}{4}$ as the usual definition for large lattices \cite{Richter1991}.

\begin{figure}
	\centering
	\includegraphics[width=0.49\textwidth]{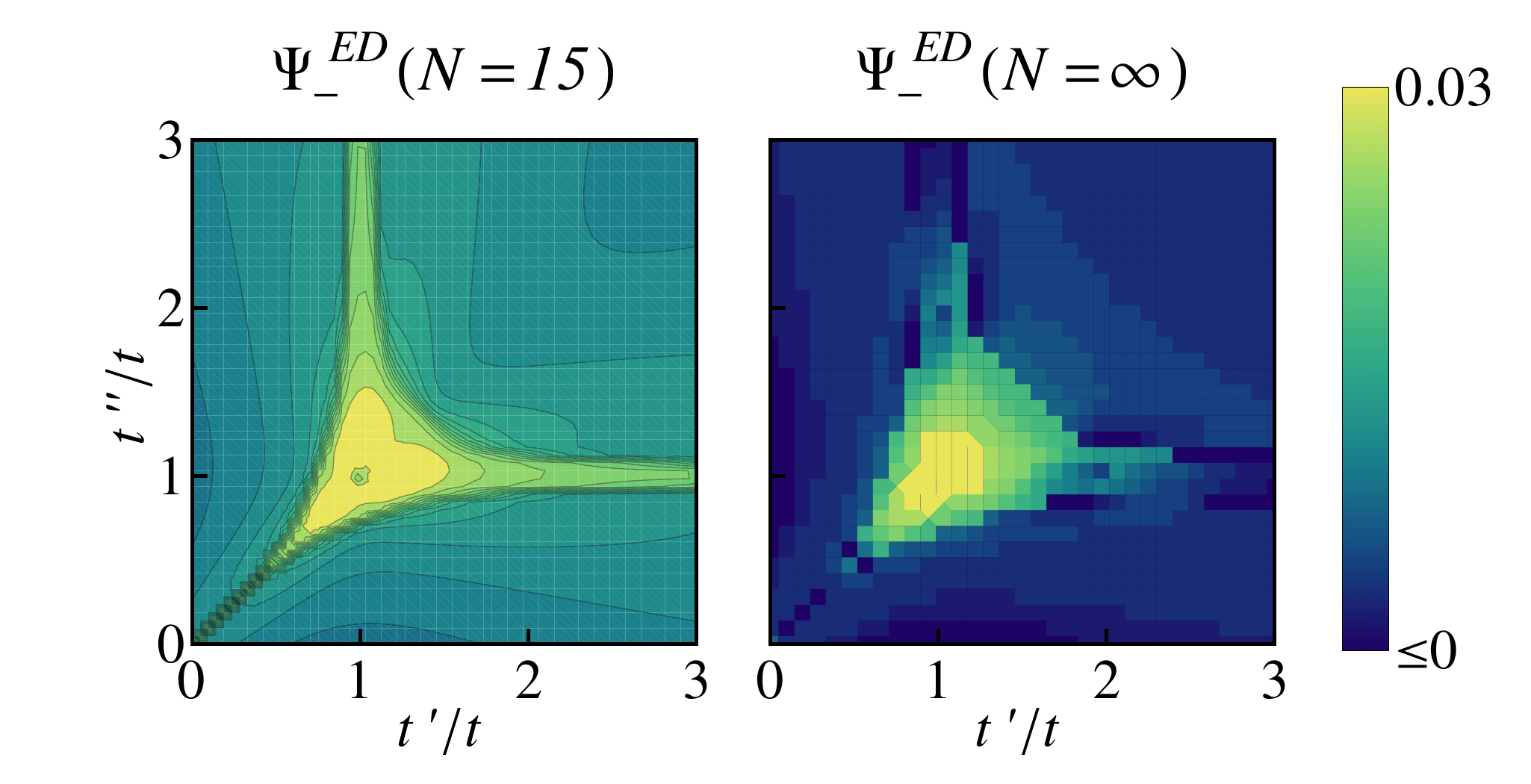}\\
	\hspace*{1.75cm}\includegraphics[width=0.4\textwidth]{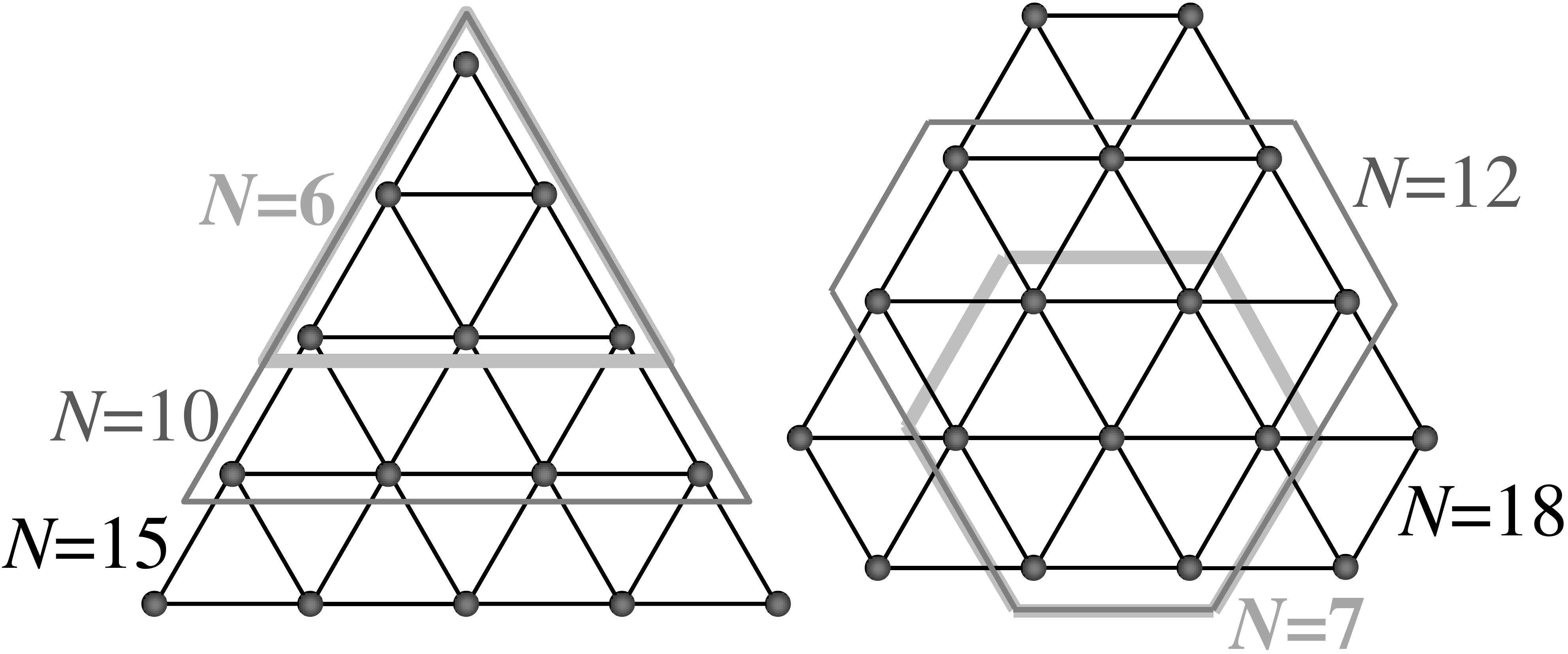}
	\caption{
	  {\bf Chiral correlations from ED.} {\bf Upper left:} Already for small systems ($N=15$), the chiral correlations are appreciably smaller in the N\'eel phases than in the rest of the phase diagram. {\bf Upper right:} From an extrapolation to large lattices, it appears that chiral LRO only survives in a small central region around $(t'/t,t''/t)=(1,1)$, lending support to the appearance of an extended disordered phase. 
{\bf Below:} The geometries used in the extrapolation are chosen for symmetry upon rotation by $60^\circ$ and equal number of $t$, $t'$, and $t''$ bonds. 
	  \label{fig:phd_ED_CC}
	}
\end{figure}

As can be seen from the ED results of the $N=15$ lattice (Fig.~\ref{fig:phd_ED_CC}, left panel), the chiral correlations are relatively small in the N\'eel phases and large in the spiral phase around $(t'/t,t''/t)=(1,1)$. However, at this lattice size, there are still appreciable chiral correlations in the rest of the parameter regime. In particular, in the 1D limit, the chiral correlations are only a little smaller than in the spiral phase. Therefore, we also plot in Fig.~\ref{fig:phd_ED_CC}, right panel, an extrapolation to large lattices by $\Psi_{-}(N)=\Psi_{-}(N=\infty)+\frac{c_1}{N^{3/2}}+\frac{c_2}{N^2}$, where we use the known form for the leading finite-size behavior \cite{Momoi1994} but also include the first subleading correction due to the small systems under consideration (our data comes from lattices with $N=7,10,12,15,18$, all chosen to have the same number of $t$, $t'$, and $t''$ bonds, as sketched at the bottom of Fig.~\ref{fig:phd_ED_CC}). 
From this it seems that the chiral correlations disappear only in the decoupled chains limits. 
Contrary to the equivalent model with Heisenberg interactions~\cite{Hauke2012a}, chiral LRO seems to not only survive in the central region around $(t'/t,t''/t)=(1,1)$, but also in part of the presumably gapped and non-magnetic region, extending all the way to the N\'eel phases. 
A finite gap coexisting with long-range chirality would indeed constitute an intriguing instance of a non-magnetic many-body ground state.

\subsection{MSWT predictions for time-of-flight pictures}

Finally, we wish to connect our predictions to experiment. A well-established experimental technique for ultracold atoms is time-of-flight (ToF) imaging of the atom momentum distribution, 
\begin{equation}
   \label{nk}
 n_b\left(\vect{k}\right)= \frac{1}{N} \sum_{i,j}e^{i \vect{k} \cdot(\vect{r}_{i}-\vect{r}_j)}  \braket{b_i^{\dagger} b_j}\,.
\end{equation}
Since the bosons of Hamiltonian~\eqref{eq:BH} have $U\to\infty$, this is equivalent to measuring the magnetic structure factor $S\left(\vect{k}\right)$, for the XY spins of Hamiltonian~\eqref{eq:HS}, see Eq.~\eqref{eq:structureFactor}.

\begin{figure*}
	\centering
	\includegraphics[width=0.75\textwidth]{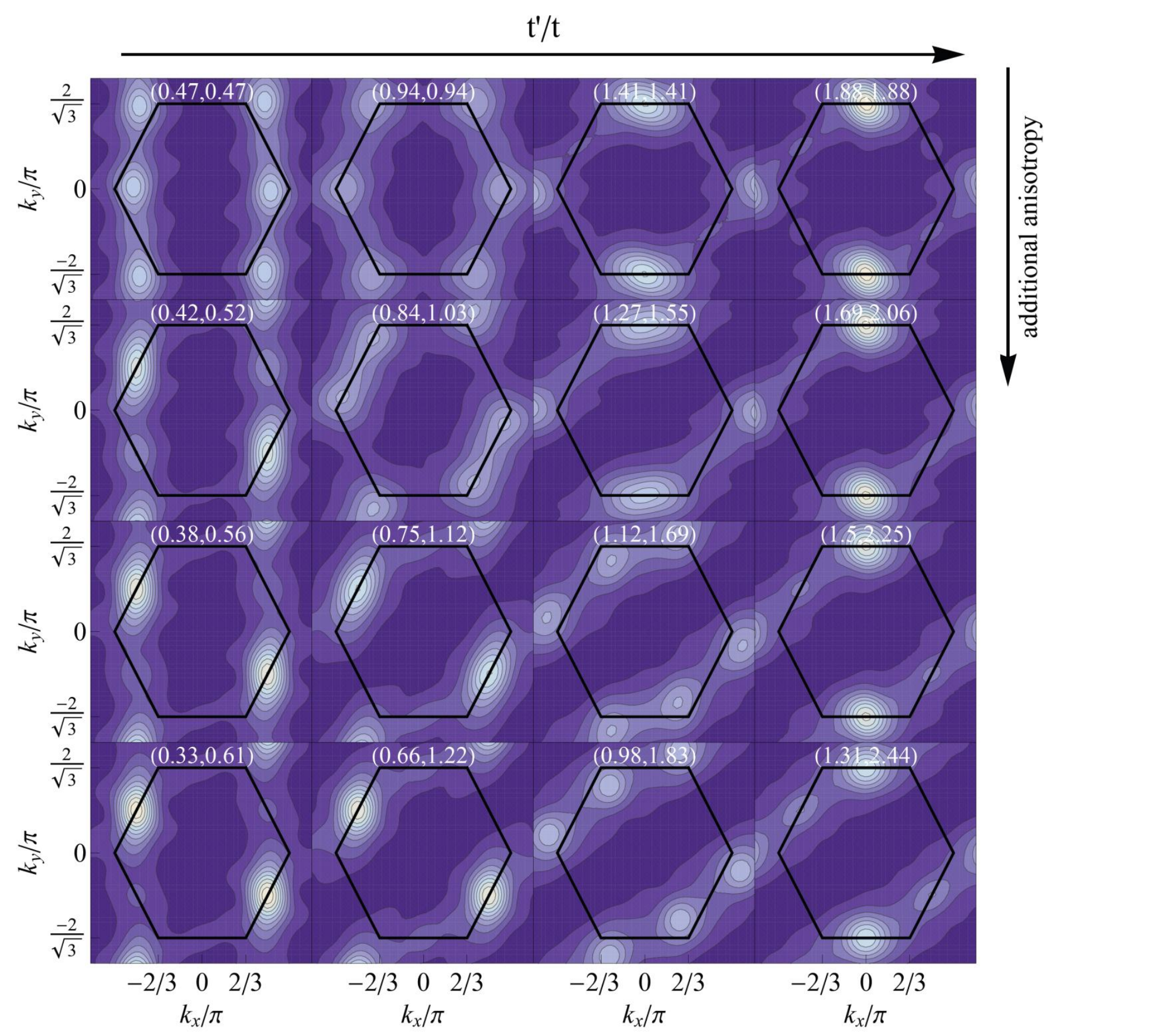}\hspace*{-2cm}\includegraphics[width=0.075\textwidth]{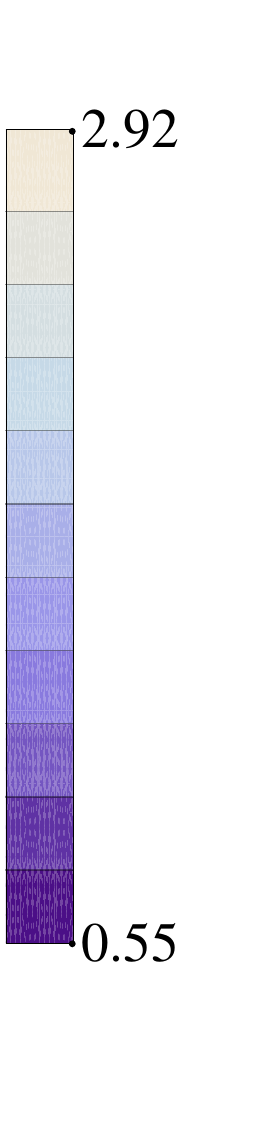}
	\caption{
	{\bf{ToF predictions}} (corresponding to the static structure factor) allow comparison to typical experimental results. 
	The additional anisotropy of the SCATL increases from top to bottom in steps of 10\%, showing that it can shift the system to phases with different qualitative order.
	\label{fig:tof}
	}
\end{figure*}

Figure~\ref{fig:tof} presents ED predictions (for $N=15$) for ToF images at various values of anisotropy. 
The uppermost row shows parameter values from the SATL. The other rows show from top to bottom results for increasing additional anisotropies in steps of 10\%. 
The black lines denote the first Brillouin  zone (1st BZ). 
Commensurate $120^\circ$ spiral order has peaks at the corners of the 1st BZ (as in the first panel of the second row), while peaks at the center of two opposing sides of the 1st BZ mark N\'eel order (as in the lower two panels of the second row). 
Incommensurate spiral order is characterized by peaks lying between these two limiting cases (as in the second panel of the second row). 
Close to the 1D limit, the peaks decrease in magnitude and smear out strongly along a straight line (as seen in the first panel of the first row). 
In large systems, disordered phases are characterized by a sub-extensive growth of the peak height.

As seen in the ToF pictures in Fig.~\ref{fig:tof}, the additional anisotropy can clearly shift the system from one phase to a qualitatively different one. For example, the momentum distributions in the first row pass from an almost 1D-like spiral state to an adjacent N\'eel phase. Similar behavior is found for other values of $(t'/t,t''/t)$. 
Such ToF pictures, therefore, would allow to observe the influence of the additional anisotropy in experiment. 

\section{Conclusion\label{cha:conclusion}}

In conclusion, we have provided a thorough analysis of the ground-state phase diagram of the quantum XY SCATL. 
Using various observables from modified spin-wave theory supplemented with ordering-vector optimization, and supported by exact diagonalization data, we have found that quantum fluctuations stabilize N\'eel order with respect to the classical phase diagram. Further, they reduce the extent of the spiral phase, which seems to be entirely surrounded by a quantum disordered region. 
This result, which constitutes our main finding, is supported by the breakdown of MSWT, together with the strong decrease of the order parameter and the spin stiffness. 
While MSWT cannot be applied to studying this region, the fact that \emph{no} semiclassical reference state describable by an ordering vector yields a stable solution is highly suggestive of a magnetically-disordered phase of purely quantum origin. Hence, our results outline a very promising candidate region for such exotic states, meriting further research with more sophisticated theoretical methods or experimental setups. 

The possible existence of quantum-disordered phases is further corroborated by ED data, where a finite gap makes magnetic LRO seem unlikely. Also, the strong decrease of the ED structure-factor peak appears to support this interpretation. 
For part of the possibly gapped region, ED suggests the persistence of chiral LRO, contrary to the equivalent model with Heisenberg interactions~\cite{Hauke2012a}. The possibility of a gapped chiral phase in this part of the phase diagram would be an intriguing subject for further research. 

A complete encircling of the spiral phase by disordered phases could naturally explain the succession of a gapped and a gapless non-magnetic phase at the low-$\alpha$ limit of the SATL. The gapless non-magnetic region would be continuously connected to the limit of decoupled chains, while the additional anisotropy of the SCATL would adiabatically connect the gapped quantum-disordered phases at small and large $\alpha$. 
Therefore, the additional anisotropy has great potential to deliver new insights into the nature of these phases. 
From an experimental point of view, the persistence of the non-magnetic phase under the additional anisotropy suggests that it can also be observed under slightly imperfect driving (i.e., if it is not completely symmetric under exchange of two bonds).

Finally, we provided MSWT predictions for the boson momentum distribution in time-of-flight pictures, a comparison to which might allow to probe the additional anisotropy in upcoming experiments, e.g., with ultracold atoms in optical lattices.\\

\textbf{Acknowledgments}
I gratefully acknowledge fruitful discussions with Tommaso Roscilde, Roman Schmied, and Luca Tagliacozzo. Also, I would like to thank Ben Powell for drawing my attention to the SCATL model. 
This work has been supported by the Catalunya Caixa, 
Spanish MICINN (FIS2008-00784), AAII-Hubbard, EU Project AQUTE, 
the Austrian Science Fund through SFB F40 FOQUS, the DARPA OLE program, 
and ERC Grant QUAGATUA.

\appendix

\section{MSWT formalism\label{cha:appendixMSWT}}

In this Appendix, we shortly review MSWT for XY
antiferromagnets; for a full description, see~\cite{Hauke2010}.

A fundamental assumption of spin-wave theory is that the ground state has LRO with ordering vector $\vect{Q}$. Hence, it is convenient to rotate the local reference system as
\begin{subequations}
\label{rot}
\begin{eqnarray}
S_i^{\,x}  &=& - \sin\left(\vect{Q}\cdot\vect{r}_i\right) S_i^{\,\eta} + \cos\left(\vect{Q}\cdot\vect{r}_i\right) S_i^{\,\zeta}\,,\\
    S_i^{\,y} &=& \phantom{-}\cos\left(\vect{Q}\cdot\vect{r}_i\right) S_i^{\,\eta} + \sin\left(\vect{Q}\cdot\vect{r}_i\right) S_i^{\,\zeta}\,,\\
    S_i^{\,z} &=& - S_i^{\,\xi}\,.
\end{eqnarray}
\end{subequations}
Then $S_i^{\,\zeta}$, which will be the quantization axis, lies parallel to the classical spin $\vect{S}_i=\left(\cos\left(\vect{Q}\cdot\vect{r}_i\right),\sin\left(\vect{Q}\cdot\vect{r}_i\right),0\right)$. This defines the classical reference state. 
We do not make any assumption on the ordering vector $\bm Q$. In particular, it may well differ from the one of the classical limit ($\vect{Q}^{\rm cl}$). 

Spin waves around this reference state can be described by the Dyson--Maleev (DM) transformation \cite{Dyson1956,Maleev1957}, which maps the physical spins to interacting bosons, 
\begin{subequations}
\label{DM}
\begin{eqnarray}
    S_i^{\,-} &\to& \frac{1}{\sqrt{2 S}}\left(2 S - a_i^\dagger a_i\right)a_i\,,\\
    S_i^{\,+} &\to& \sqrt{2 S}\, a_i^\dagger,\\
    S_i^{\,\zeta} &\to& -S+a_i^\dagger a_i\,,
\end{eqnarray}
\end{subequations}
where $S_i^{\,\pm}\equiv S_i^{\,\xi}\pm i S_i^{\,\eta}$. These bosons describe quantum fluctuations around the spin reference state, and are not to be confused with the original bosons appearing in Hamiltonian~\eqref{eq:BH}.
 
\begin{widetext}
This transformation maps the Hamiltonian, Eq.~\eqref{eq:HS}, to the non-linear bosonic Hamiltonian
\begin{align}
\label{H4}
   {\cal H}&= S^2 \sum_{\braket{i,j}} t_{ij} \left[ 
    1 - \frac{1}{2S} \left(2 a_i^\dagger a_i + 2 a_j^\dagger a_j-a_i^\dagger a_j - a_i a_j^\dagger +a_i^\dagger a_j^\dagger+ a_i a_j\right)  \right.  \nonumber \\
    & 
    \left. + \frac{1}{(2S)^2} 
    \left( a_i^\dagger a_i a_j^\dagger a_j - a_i^\dagger a_j^\dagger a_j a_j - a_i^\dagger a_i a_i a_j^\dagger + a_i a_j^\dagger a_j a_j + a_i^\dagger a_i a_i a_j \right)
     + {\cal O}\left(\frac{n}{2S}\right)^3 \right]  \cos\left(\vect{Q}\cdot\vect{r}_{ij}\right)\,. 
\end{align}
\end{widetext}
where $a_i$ ($a_i^{\dagger}$) destroys (creates) a DM boson at site $i$, and $S$ is the length of the spin.
Here, we neglected the kinematic constraint which restricts the DM-boson density $n$ to the physical subspace $n < 2S$. Moreover, we dropped terms with six boson operators, which are of order  ${\cal O}[n/(2S)^3]$ and are negligible for $n/(2S)<1$. 
Using Wick's theorem \cite{Fetter1971}, and defining the correlators $\braket{a_i^\dagger a_j}=F\left( \vect{r}_{ij} \right)-\frac 1 2 \delta_{ij}$ and $\braket{a_i a_j}= \braket{a_i^\dagger a_j^\dagger} \,\, = \,\, G\left( \vect{r}_{ij} \right)$, the expectation value $E\equiv\braket{\cal H}$ can be written as
\begin{eqnarray}
    E &= &   \frac 1 2 \sum_{\braket{i,j}} t_{ij} \left\lbrace \left[S+\frac 1 2 - F\left( 0 \right) + F\left( \vect{r}_{ij} \right) \right]^2 \right.  \\
    	&  & + \left. \left[S+\frac 1 2 - F\left( 0 \right) + G\left( \vect{r}_{ij} \right) \right]^2 \,\right\rbrace \cos\left(\vect{Q}\cdot\vect{r}_{ij}\right) \,. \nonumber
\end{eqnarray}
After Fourier transforming, $a_{\vect{k}}=\frac{1}{\sqrt{N}}\sum_i a_i\, \ue^{-i \vect{k}\cdot\vect{r}_i}$, 
and a subsequent Bogoliubov transformation, $\alpha_{\vect{k}\phantom{-}} = \phantom{-}\cosh\theta_{\vect{k}}\, a_{\vect{k}} - \sinh\theta_{\vect{k}} \, a_{-\vect{k}}^\dagger$, and $\alpha_{-\vect{k}}^\dagger = -\sinh\theta_{\vect{k}} \, a_{\vect{k}} + \cosh\theta_{\vect{k}} \, a_{-\vect{k}}^\dagger$, we 
minimize the free energy $\mathcal{F}$ under the constraint of vanishing magnetization at each site, $\braket{a_i^{\dagger} a_i} = S$, which is known as Takahashi's modification \cite{Takahashi1989}. 
This yields a set of self-consistent equations, 
\begin{equation}
    \label{tanh2th}
    \tanh 2\theta_{\vect{k}}=\frac{A_{\vect{k}}}{B_{\vect{k}}}
\end{equation}
with
\begin{subequations}
    \label{AkBk}
\begin{eqnarray}
    \label{Ak}
    A_{\vect{k}} & = & -\frac 1 N \sum_{\braket{i,j}} t_{ij} \cos\left({\vect{Q}\cdot\vect{r}_{ij}}\right) G_{ij} \,\ue^{i\vect{k}\cdot\vect{r}_{ij}}\,,\\
    \label{Bk}
    B_{\vect{k}} &= & - \frac 1 N \sum_{\braket{i,j}} \left\lbrace t_{ij} \cos\left({\vect{Q}\cdot\vect{r}_{ij}}\right) \right. \nonumber \\
    & & \qquad \left. \times \left[ G_{ij} + F_{ij} \left(1-\ue^{i\vect{k}\cdot\vect{r}_{ij}}\right)\right] \right\rbrace - \mu  
\end{eqnarray}
\end{subequations}
where $\mu$ is the Lagrange multiplier for Takahashi's constraint.
The spin-wave spectrum reads
\begin{equation}
\label{disp}
\omega_{\vect{k}}=\sqrt{B_{\vect{k}}^2-A_{\vect{k}}^2}\,.
\end{equation}
At $T=0$, $\mu$ vanishes, which implies the disappearance of the gap at $\vect{k}=0$ that may exist for finite temperature. This is a necessary condition for magnetic LRO, and enables Bose condensation in the $\vect{k}=0$ mode. 
Separating out its contribution, 
\eq{
\braket{a_{\vect{k}=0}^{\dagger} a_{\vect{k}=0}}/N=\braket{a_{\vect{k}=0} a_{\vect{k}=0}}/N\equiv M\,,
\label{eq:orderParameterMSWT}
}
which corresponds to the magnetic order parameter, 
one arrives at the zero-temperature equations
\begin{subequations}
\label{FG}
\begin{eqnarray}
    F_{ij}&=&M + \frac 1 {2 N} \sum_{\vect{k}\neq 0} \frac{B_{\vect{k}}} {\omega_{\vect{k}}}\cos\left(\vect{k}\cdot\vect{r}_{ij}\right)\label{Fij}\,,\\
    G_{ij}&=&M + \frac 1 {2 N} \sum_{\vect{k}\neq 0} \frac{A_{\vect{k}}} {\omega_{\vect{k}}}\cos\left(\vect{k}\cdot\vect{r}_{ij}\right)\label{Gij}\,,
\end{eqnarray}
\end{subequations}
and the constraint of vanishing magnetization at each site becomes
\begin{equation}
    \label{constr2}
    S+\frac 1 2 = M + \frac 1 {2 N} \sum_{\vect{k}\neq 0} \frac{B_{\vect{k}}} {\omega_{\vect{k}}}.
\end{equation}

It is not \emph{a priori} clear that the classical ordering vector $\vect{Q}^{\mathrm{cl}}$ correctly describes the LRO in the quantum system. 
To account for a competition between LRO at different ordering vectors $\vect{Q}$, we extend the MSWT procedure by optimizing the free energy $\mathcal{F}$ with respect to the ordering vector $\vect{Q}$.
This yields two additional equations which must be added to the set of self-consistent equations, 
\begin{subequations}
\label{Qs}
\begin{equation}
    \label{Qx}
    \frac\partial{\partial Q_x} \mathcal{F}=-\frac 1 2 \sum_{\braket{i,j}} t_{ij} \sin\left(\vect{Q}\cdot\vect{r}_{ij}\right)r_{ij}^x\left[F_{ij}^2+G_{ij}^2\right] = 0 \,,
\end{equation}
\begin{equation}
    \label{Qy}
    \frac\partial{\partial Q_y} \mathcal{F}=-\frac 1 2 \sum_{\braket{i,j}} t_{ij} \sin\left(\vect{Q}\cdot\vect{r}_{ij}\right)r_{ij}^y\left[F_{ij}^2+G_{ij}^2\right]=0\,.
\end{equation}
\end{subequations}

The values of $F_{ij}$ and $G_{ij}$ can now be calculated by solving self-consistently Eqs.~(\ref{AkBk}--\ref{Qs}).
Through Wick's theorem the knowledge of the quantities $F_{ij}$ and $G_{ij}$ 
determines the expectation value of any observable.


\begin{thebibliography}{10}

\bibitem{Lewenstein2012}
Lewenstein, M., Sanpera, A., and Ahufinger, V.
\newblock {\em Ultracold Atoms in Optical Lattices: Simulating Quantum
  Many-Body Systems}.
\newblock Oxford University Press, Oxford,  (2012).

\bibitem{Jaksch2003}
Jaksch, D. and Zoller, P.
\newblock {\em New J. Phys.}{ \bf 5}, 56 (2003).

\bibitem{Sorensen2005}
S\o{}rensen, A.~S., Demler, E., and Lukin, M.~D.
\newblock {\em Phys. Rev. Lett.}{ \bf 94}, 086803 (2005).

\bibitem{Lin2009b}
Lin, Y.-J., Compton, R.~L., Jim\'{e}nez-Garc\'{i}a, K., Porto, J.~V., and
  Spielman, I.~B.
\newblock {\em Nature}{ \bf 462}, 628 (2009).

\bibitem{Aidelsburger2011}
Aidelsburger, M., Atala, M., Nascimb\`ene, S., Trotzky, S., Chen, Y.-A., and
  Bloch, I.
\newblock {\em Phys. Rev. Lett.}{ \bf 107}, 255301 (2011).

\bibitem{Jimenez2012}
Jim{\'i}nez-Garc{\'i}a, K., LeBlanc, L.~J., Williams, R.~A., Beeler, M.~C.,
  Perry, A.~R., and Spielman, I.~B.
\newblock {\em Phys Rev Lett.} { \bf 108}, 225303 (2012).

\bibitem{Eckardt2010}
Eckardt, A., Hauke, P., Soltan-Panahi, P., Becker, C., Sengstock, K., and
  Lewenstein, M.
\newblock {\em Europhys. Lett.}{ \bf 89}, 10010 (2010).

\bibitem{Struck2011a}
Struck, J., \"{O}lschl\"{a}ger, C., {Le Targat}, R., Soltan-Panahi, P.,
  Eckardt, A., Lewenstein, M., Windpassinger, P., and Sengstock, K.
\newblock {\em Science}{ \bf 333}, 996 (2011).

\bibitem{Sacha2011}
Sacha, K., Targonska, K., and Zakrzewski, J.
\newblock {\em Phys. Rev. A}{ \bf 85}, 053613 (2012).

\bibitem{Struck2012}
Struck, J., \"Olschl\"ager, C., Weinberg, M., Hauke, P., Simonet, J., Eckardt,
  A., Lewenstein, M., Sengstock, K., and Windpassinger, P.
\newblock {\em Phys. Rev. Lett.}{ \bf 108}, 225304 (2012).

\bibitem{Hauke2012c}
Hauke, P., Tieleman, O., Celi, A., \"Olschl\"ager, C., Simonet, J., Struck, J., Weinberg, M., Windpassinger, P.,  Sengstock, K., Lewenstein, M., and Eckardt, A.
\newblock {\em Phys. Rev. Lett.}{ \bf 109}, 145301 (2012).

\bibitem{Goldbaum2008}
Goldbaum, D.~S. and Mueller, E.~J.
\newblock {\em Phys. Rev. A}{ \bf 77}, 033629 (2008).

\bibitem{Garcia-Ripoll2007}
Garcia-Ripoll, J.~J. and Pachos, J.~K.
\newblock {\em New J. Phys.}{ \bf 9}, 139 (2007).

\bibitem{Diep2004}
Diep, H.~T., editor.
\newblock {\em Frustrated Spin Systems}.
\newblock World Scientific, Singapore,  (2004).

\bibitem{Schmied2008}
Schmied, R., Roscilde, T., Murg, V., Porras, D., and Cirac, J.~I.
\newblock {\em New J. Phys.}{ \bf 10}, 045017 (2008).

\bibitem{Fazio2001}
Fazio, R. and van~der Zant, H.
\newblock {\em Phys. Rep.}{ \bf 355}, 235 (2001).

\bibitem{Hauke2010}
Hauke, P., Roscilde, T., Murg, V., Cirac, J.~I., and Schmied, R.
\newblock {\em New J. Phys.}{ \bf 12}, 053036 (2010).

\bibitem{Friedenauer2008}
Friedenauer, A., Schmitz, H., Glueckert, J.~T., Porras, D., and Schaetz, T.
\newblock {\em Nat. Phys.}{ \bf 4}, 757 (2008).

\bibitem{Islam2011}
Islam, R., Edwards, E., Kim, K., Korenblit, S., Noh, C., Carmichael, H., Lin,
  G.-D., Duan, L.-M., Wang, C.-C.~J., Freericks, J., and Monroe, C.
\newblock {\em Nature Communications}{ \bf 2}, 377 (2011).

\bibitem{Kim2010}
Kim, K., Chang, M.-S., Korenblit, S., Islam, R., Edwards, E.~E., Freericks,
  J.~K., Lin, G.-D., Duan, L.-M., and Monroe, C.
\newblock {\em Nature}{ \bf 465}, 590 (2010).

\bibitem{Takahashi1989}
Takahashi, M.
\newblock {\em Phys. Rev. B}{ \bf 40}, 2494 (1989).

\bibitem{Hauke2011}
Hauke, P., Roscilde, T., Murg, V., Cirac, J., and Schmied, R.
\newblock {\em New J. Phys.}{ \bf 13}, 075017 (2011).

\bibitem{Hauke2012a}
Hauke, P. 
\newblock {\em Phys. Rev. B}{ \bf 87}, 014415 (2013);
\newblock see also preceding article {\em arXiv:1205.1955 [cond-mat.str-el]}{ \bf } (2012).

\bibitem{Sandvik1999}
Sandvik, A.~W. and Hamer, C.~J.
\newblock {\em Phys. Rev. B}{ \bf 60}, 6588 (1999).

\bibitem{Momoi1994}
Momoi, T.
\newblock {\em J. Stat. Phys.}{ \bf 75}, 707 (1994).




\bibitem{Chandra1988}
Chandra, P. and Doucot, B.
\newblock {\em Phys. Rev. B}{ \bf 38}, 9335 (1988).

\bibitem{Locher1990}
Locher, P.
\newblock {\em Phys. Rev. B}{ \bf 41}, 2537 (1990).

\bibitem{Ferrer1993}
Ferrer, J.
\newblock {\em Phys. Rev. B}{ \bf 47}, 8769 (1993).

\bibitem{Zhong1993}
Zhong, Q.~F. and Sorella, S.
\newblock {\em Europhys. Lett.}{ \bf 21}, 629 (1993).

\bibitem{Leung1996}
Leung, P.~W. and Lam, N.
\newblock {\em Phys. Rev. B}{ \bf 53}, 2213 (1996).

\bibitem{Capriotti2004a}
Capriotti, L., Scalapino, D.~J., and White, S.~R.
\newblock {\em Phys. Rev. Lett.}{ \bf 93}, 177004 (2004).

\bibitem{Capriotti2004b}
Capriotti, L. and Sachdev, S.
\newblock {\em Phys. Rev. Lett.}{ \bf 93}, 257206 (2004).

\bibitem{Mambrini2006}
Mambrini, M., L{\"a}uchli, A., Poilblanc, D., and Mila, F.
\newblock {\em Phys. Rev. B}{ \bf 74}, 144422 (2006).

\bibitem{Shannon2006}
Shannon N., Momoi T., and Sindzingre P.
\newblock {\em Phys. Rev. Lett.}{ \bf 96}, 027213 (2006).

\bibitem{Murg2009}
Murg, V., Verstraete, F., and Cirac, J.~I.
\newblock {\em Phys. Rev. B}{ \bf 79}, 195119 (2009).

\bibitem{Richter2010}
Schulenburg, J. and Richter, J.
\newblock {\em Eur. Phys. J. B}{ \bf 73}, 117 (2010).

\bibitem{Reuther2011b}
Reuther J., W{\"o}lfle P., Darradi R., Brenig W., Arlego M., and Richter J.
\newblock {\em Phys. Rev. B}{ \bf 83}, 064416 (2011).


\bibitem{Farnell2011}
Farnell D.~J.~J., Bishop R.~F., Li P.~H.~Y., Richter J., and Campbell C.~E. 
\newblock {\em Phys. Rev. B}{ \bf 84}, 012403 (2011).

\bibitem{Albuquerque2011}
Albuquerque A.~F., Schwandt D., Het\'enyi B., Capponi S., Mambrini M., and L\"auchli A.~M.
\newblock {\em Phys. Rev. B}{ \bf 84}, 024406 (2011).

\bibitem{Li2012}
Li P.~H.~Y., Bishop R.~F., Farnell D.~J.~J., Richter J., and Campbell C.~E. 
\newblock {\em Phys. Rev. B}{ \bf 85}, 085115 (2012).

\bibitem{Varney2011}
Varney C.~N., Sun K., Galitski V., and Rigol M.
\newblock {\em Phys. Rev. Lett.}{ \bf 107}, 077201 (2011).




\bibitem{Stephenson1969}
Stephenson, J.
\newblock {\em Can. J. Phys.}{ \bf 47}, 2621 (1969).

\bibitem{Stephenson1970}
Stephenson, J.
\newblock {\em Can. J. Phys.}{ \bf 48}, 1724 (1970).

\bibitem{Stephenson1970a}
Stephenson, J.
\newblock {\em Can. J. Phys.}{ \bf 48}, 2118 (1970).

\bibitem{Stephenson1970b}
Stephenson, J.
\newblock {\em J. Math. Phys.}{ \bf 11}, 420 (1970).

\bibitem{Chernyshev2009}
Chernyshev, A.~L. and Zhitomirsky, M.~E.
\newblock {\em Phys. Rev. B}{ \bf 79}, 144416 (2009).

\bibitem{Chubukov1994}
Chubukov, A.~V., Sachdev, S., and Ye, J.
\newblock {\em Phys. Rev. B}{ \bf 49}, 11919 (1994).

\bibitem{Halperin1969}
Halperin, B.~I. and Hohenberg, P.~C.
\newblock {\em Phys. Rev.}{ \bf 188}, 898 (1969).

\bibitem{Lhuillier2005}
Lhuillier, C.
\newblock {\em arXiv:cond-mat/0502464v1}{ \bf } (2005).

\bibitem{Richter1991}
Richter, J., Gros, C., and Weber, W.
\newblock {\em Phys. Rev. B}{ \bf 44}, 906 (1991).

\bibitem{Dyson1956}
Dyson, F.~J.
\newblock {\em Phys. Rev.}{ \bf 102}, 1217 (1956).

\bibitem{Maleev1957}
Maleev, S.~V.
\newblock {\em Zh. Eksp. Teor. Fiz.}{ \bf 30}, 1010 (1957).
\newblock see also Sov. Phys. JETP 6, 776 (1958).

\bibitem{Fetter1971}
Fetter, A. and Walecka, J.
\newblock {\em Quantum Theory of Many-Particle Systems}.
\newblock McGraw Hill, New York,  (1971).

\end{thebibliography}
\end{document}